\documentclass[aps,prc,onecolumn,preprint,nofootinbib]{revtex4-2}
\usepackage{amsmath,amssymb} 
\usepackage{physics}
\usepackage{here} 
\usepackage[dvipdfmx]{graphicx} 

\begin{document}

\title{Pairing effects on pure rotational energy of nuclei}

\author{K.~Abe}
\affiliation{Department~of~Physics, Graduate~School~of~Science~and~Engineering,
  Chiba~University, Yayoi-cho 1-33, Inage, Chiba~263-8522, Japan}

\author{H.~Nakada}
  \email{nakada@faculty.chiba-u.jp}
\affiliation{Department~of~Physics, Graduate~School~of~Science,
  Chiba~University, Yayoi-cho 1-33, Inage, Chiba~263-8522, Japan}

\begin{abstract}
  By applying the angular-momentum projection (AMP)
  to the self-consistent axial mean-field solutions
  with the semi-realistic effective Hamiltonian M3Y-P6,
  the pairing effects on the pure rotational energy of nuclei,
  \textit{i.e.}, the rotational energy at a fixed intrinsic state,
  have been investigated.
  While it was shown at the Hartree-Fock (HF) level
  that the individual terms of the Hamiltonian
  contribute to the rotational energy with ratios insensitive to nuclides
  except for light or weakly-deformed nuclei,
  the pair correlations significantly change the contributions,
  even for the well-deformed heavy nuclei.
  The contribution of the interaction to the rotational energy
  is found to correlate well with the degree of proximity between nucleons,
  which is measured via the expectation value
  that two nucleons exist at the same position.
  While the nucleons slightly spread as the angular momentum increases
  at the HF level,
  accounting for the positive (negative) contribution
  of the attractive (repulsive) components of the interaction,
  the pair correlations reduce or invert the effect.
\end{abstract}

\maketitle

\section{Introduction}

It is known that the rotational band appears in a number of nuclei,
giving the excitation energy $E_{\mathrm{x}}(J^+)=J(J+1)/(2\mathcal{I})$
above the ground state (g.s.)~\cite{BM98,NNDC}.
These energy spectra indicate that the intrinsic state of the nucleus
is deformed and rotates with the moment of inertia (MoI) $\mathcal{I}$.
From a microscopic standpoint,
nuclei are well described in the self-consistent mean-field (MF) theories,
such as the Hartree-Fock (HF)
and the Hartree-Fock-Bogoliubov (HFB) approximations~\cite{RS80},
and the deformed intrinsic states of nuclei are obtained as MF solutions
in which the rotational symmetry is spontaneously broken.
The Nambu-Goldstone (NG) mode is accompanied by symmetry breaking,
and it restores the corresponding symmetry in energy eigenstates.
The restoration of the rotational symmetry corresponds
with a whole rotation of a deformed intrinsic state.
The superposition of the degenerate intrinsic states along the NG mode
derives the angular-momentum projection (AMP)~\cite{RS80,PY57,Yo57,PT62,
  Ve63,Ve64,OY66,Ka68,BB69,Ma75,HI79,HIT79,HI80,RER02,BHR03,Sc04,BH08,RRG19,
  BB21,SDRRY21,AN22}.
The AMP accounts for the $J(J+1)$ rule of the excitation energy
under a reasonable approximation
for well-deformed heavy nuclei~\cite{RS80,PY57,Yo57,Ve63,Ve64,Ka68}.
However, rotational spectra have also been observed in light nuclei
including those far off the $\beta$-stability.
It will deserve reinvestigating from a general perspective
how the rotational energy of nuclei arises,
not restricting ourselves to well-deformed heavy nuclei.

In classical mechanics,
the rotational energy arises from the kinetic energy,
proportional to $\mathbf{J}^2/(2\mathcal{I})$ for a rigid rotor
with the angular momentum $\mathbf{J}$,
in which the distance between constituent particles is invariant.
On the other hand,
the above-mentioned rotational energy of nuclei should be formed
from the effective Hamiltonian, including the nucleonic interaction.
In the Thouless-Valatin formula for the MoI~\cite{TV62},
which is derived from the random-phase approximation (RPA),
the g.s. correlations due to the interaction are taken into account.
In Ref.~\cite{AN22},
we investigated the composition of the rotational energy of various nuclei
by applying the AMP to the HF solutions.
When the rotational energy is decomposed
into contributions of the individual terms of the Hamiltonian,
those of the central forces are sizable,
although the kinetic energies carry the major part
and are close to the rigid-rotor values.
The ratios of the individual terms of the Hamiltonian
to the total rotational energy are insensitive to nuclides and deformation,
as long as they are well-deformed and not very light.
In contrast,
the ratios significantly depend on nuclei and deformation
for light or weakly-deformed nuclei.

It is known that the pair correlations in nuclei
significantly influence the rotational energy.
Within the cranking model~\cite{RS80},
the Belyaev formula~\cite{Be59} accounts for the reduction of the MoI
compared with the Inglis formula~\cite{In5455}
due to the pair correlations.
Conversely, the pair correlations could be influenced by the rotation;
the pair correlation is suppressed by the cranking effect
accompanied by the breaking of the time-reversal symmetry,
known as the Mottelson-Valatin effect~\cite{MV60}.

The effects of the pair correlations on the rotational energy of nuclei
are investigated in this paper,
extending the study in Ref.~\cite{AN22}.
The AMP is applied to the self-consistent solutions
of the Bardeen-Cooper-Schrieffer approximation on top of the axial HF orbitals
(HF+BCS)
as well as the axial HFB solutions.
The Michigan-three-range-Yukawa (M3Y)-type nucleonic interactions~\cite{
  Na03,Na13,Na20}
have been developed for the self-consistent MF calculations.
With phenomenological modification of the original M3Y interaction
obtained from the $G$-matrix calculations~\cite{BBML77,CLL73,LLR80,ATB83},
they may be regarded as \textit{semi-realistic} effective interactions.
They have been pointed out to be free from most of the instabilities
in the nuclear matter response functions,
which sometimes occur in other interactions~\cite{DPN21}.
The M3Y-P6 parameter-set is employed~\cite{Na13,Na20},
which well describes the magic numbers of nuclei
over a wide range of the nuclear chart~\cite{NS1416}
and the deformation of nuclei~\cite{SNM16,MN18}.

In the present study,
the AMP is applied to the MF wave functions
obtained by self-consistent axial-MF calculations
with the M3Y-P6 interaction,
including the pair correlations.
We restrict ourselves to the energies
arising solely from the rotation of the MF solutions
in the projection-after-variation (PAV) framework as in Ref.~\cite{AN22},
which may be called \textit{pure rotational energy}\footnote{
  Since the state is restricted to the axial HF state as in Ref.~\cite{PY57},
  the corresponding energy was called \textit{Peierls-Yoccoz rotational energy}
  in Ref.~\cite{AN22}.},
separating them from the counter effects of the rotation
on the intrinsic states.
It should be kept in mind that the pure rotational energy
is not necessarily enough to describe the rotational spectra
in actual nuclei \cite{PT62},
because the rotation may significantly affect the intrinsic state
as handled in the cranking model \cite{PT62,Ka68,RS80}
and the variation-after-projection (VAP) schemes \cite{RS80}.

The correlations of the spatial proximity between two neutrons (protons)
have attracted interest as di-neutron (di-proton) correlations~\cite{MKY05}.
The spatial correlations between nucleons
could be relevant to the contributions of the nucleonic interaction
to the rotational energy,
giving us an insight into why the interaction contributes.

\section{Theoretical framework and numerical method}
\label{sec:theory}

\subsection{AMP and cumulant expansion}
\label{subsec:theory}

The AMP is the method by which an intrinsic state is projected on angular-momentum eigenstates~\cite{RS80}.
We consider an axially-symmetric intrinsic state,
which is an eigenstate of $\hat{J}_z$ with the eigenvalue $M=0$.
The intrinsic state $\ket{\Phi_0}$ is expanded
by angular-momentum eigenstates $\ket{J0}$,
$\ket*{\Phi_0} = \sum_J \ket*{J0} \braket*{J0}{\Phi_0}$,
where we omit indices other than $J$ and $M\,(=0)$ for simplicity.
The Wigner (small) $d$ function~\cite{RS80,VMK,JJ94} of the angle $\beta$,
$d^{(J)}_{MK}(\beta) := \mel*{JM} {e^{-i \hat{J}_y \beta}} {JK}$,
takes a real number under the standard phase convention.
The expectation values of the scalar operator $\hat{\mathcal{S}}$
on the angular-momentum eigenstates are obtained as follows~\cite{RS80,AN22}:
\begin{equation}\label{eq:SJ}
\ev*{ \hat{\mathcal{S}} }{ J }
=\,\frac{\displaystyle \int_0^{\pi/2} d\beta \sin\beta\,d^{(J)}_{00}(\beta)
\ev*{\hat{\mathcal{S}}\,e^{-i \hat{J}_y \beta}}{\Phi_0}}
{\displaystyle \int_0^{\pi/2} d\beta \sin\beta\,d^{(J)}_{00}(\beta)
\ev*{e^{-i \hat{J}_y \beta}}{\Phi_0}}.
\end{equation}
Equation~\eqref{eq:SJ} is the basic formula of the AMP.
We here omit the index $M\,(=0)$ on the left-hand side (LHS)
of Eq.~\eqref{eq:SJ}.
The energy difference $\ev*{\hat{H}}{J} - \ev*{\hat{H}}{0}$,
where $\hat{H}$ is the Hamiltonian, is the pure rotational energy.

It has been pointed out in Ref.~\cite{AN22}
that the cumulant expansion can straightforwardly be applied
to the right-hand side (RHS) of Eq.~\eqref{eq:SJ}.
We expand $d^{(J)}_{00}(\beta)$ by the power series of $\beta$,
\begin{equation}\label{eq:Taylor_dfun_c2n}
d^{(J)}_{00}(\beta)
= \sum_{n=0}^{\infty} c_{2n} \beta^{2n};\,\quad
c_{2n} =\,\frac{(-)^n}{(2n)!} \ev*{\hat{J}_y^{\,2n}}{J0}.
\end{equation}
Note $c_0=1$, $c_2=-J(J+1)/(2!\,2)$ and $c_{2n}=\mathcal{O}(J^{2n})$.
The function $\mathcal{S}^{01}(\beta)$ is defined
and also expanded by the power series of $\beta$ as follows:
\begin{equation}\label{eq:def_S01_s2n}
\mathcal{S}^{01}(\beta)
:=\,\frac{\ev*{\hat{\mathcal{S}}\,e^{-i \hat{J}_y \beta}}{\Phi_0}}
{\ev*{e^{-i \hat{J}_y \beta}}{\Phi_0}}
= \sum_{n=0}^{\infty} s_{2n} \beta^{2n};\quad s_{2n}
=\,\frac{(-)^n} {(2n)!} \ev*{\hat{\mathcal{S}};
  \underbrace{\hat{J}_y;\cdots;\hat{J}_y}_{2n}}{\Phi_0}_{\mathrm{cum}},
\end{equation}
with the cumulant defined for a set of commutable operators
$\{\hat{X}_i; i=1,2,\cdots,n\}$~\cite{Ku62},
\begin{equation}\label{eq:def_cum}
\ev*{ \hat{X}_1;\cdots;\hat{X}_n}_{\mathrm{cum}}
:= \frac{\partial}{\partial t_1}\cdots\frac{\partial}{\partial t_n}
\left.\ln \ev{\exp\left(\sum_{i=1}^{n} t_i {\hat{X}}_i \right)}
\right|_{t_1=\cdots=t_n=0}.
\end{equation}
The low-order cumulants are represented as
\begin{equation}
s_0 = \ev*{ \hat{\mathcal{S}} }{ \Phi_0 },\quad
s_2 = -\frac{1}{2!} \, C[ \hat{\mathcal{S}}, \hat{J}_y^{\,2} ],\quad
s_4 = \,\frac{1}{4!}
\left( C[\hat{\mathcal{S}}, \hat{J}_y^{\,4}]
-6\,C[\hat{\mathcal{S}}, \hat{J}_y^{\,2}] (\sigma[\hat{J}_y])^2\right).
\end{equation}
Here $\sigma[\hat{A}]$ is the fluctuation of an operator $\hat{A}$,
and $C[\hat{A}, \hat{B}]$ is the correlation function
of operators $\hat{A}$ and $\hat{B}$,
\begin{equation}\label{eq:def_C_A-B}
C[\hat{A}, \hat{B}]:=
\ev*{\hat{A} \hat{B}}{\Phi_0}-\ev*{\hat{A}}{\Phi_0} \ev*{\hat{B}}{\Phi_0},\quad
\sigma[\hat{A}]:=\sqrt{C[\hat{A}, \hat{A}]}.
\end{equation}
Then Eq.~\eqref{eq:SJ} is exactly expressed in terms of the cumulants,
\begin{equation}\label{eq:JSJ_Lambda}
\ev*{ \hat{\mathcal{S}} }{ J }
=\,\frac{\displaystyle \sum_{m,n=0}^{\infty} c_{2m} s_{2n} \varLambda_{2m+2n}}
{\displaystyle \sum_{\ell=0}^{\infty} c_{2\ell} \varLambda_{2\ell}},
\end{equation}
where
\begin{equation}\label{eq:N_2n_Lambda_2n}
N_{2n}
:= \int_0^{\pi/2}d\beta \sin \beta\,\beta^{2n}
\ev*{e^{-i \hat{J}_y \beta}}{\Phi_0},\quad
\varLambda_{2n}
:= \frac{N_{2n}}{N_0},\quad (n=0,1,2,\cdots).
\end{equation}
Expansion of Eq.~\eqref{eq:JSJ_Lambda} with respect to $c_{2n}$
yields the g.s. expectation value and the $J(J+1)$ rule
for $\ev*{\hat{\mathcal{S}}}{J}$,
\begin{subequations}\label{eq:J(J+1)_def_moi}
\begin{align}
\ev*{\hat{\mathcal{S}}}{J} =&\,
\ev*{\hat{\mathcal{S}}}{0}
+ \frac{J(J+1)}{2\,\mathcal{I}[\mathcal{S}]}+\cdots\,;
\label{eq:J(J+1)}\\
\ev*{\hat{\mathcal{S}}}{0} :=&\,
\displaystyle \sum_{n=0}^{\infty} s_{2n} \varLambda_{2n},\quad
\frac{1}{\mathcal{I}[\hat{\mathcal{S}}]} :=\,
\sum_{n=1}^{\infty} s_{2n}
\left[-\frac{1}{2} (\varLambda_{2n+2}-\varLambda_{2n} \varLambda_2)\right].
\label{eq:def_moi}
\end{align}
\end{subequations}
The MoI of Peierls and Yoccoz~\cite{PY57,Ve63}
is obtained by neglecting the $s_{2n}$ terms with $n\geq 2$
for $\hat{\mathcal{S}}=\hat{H}$.
We shall call the approximation of Eq.~\eqref{eq:J(J+1)_def_moi}
up to the $s_2$ terms
\textit{Peierls-Yoccoz (PY) formula}.
It has been shown that the higher-order terms are not negligible
in light nuclei or weakly-deformed intrinsic states~\cite{AN22}.

The cumulant of Eq.~\eqref{eq:def_cum} can be generalized as
\begin{equation}\label{eq:def_cum2}
\ev*{ \hat{X}_1 ; \cdots ; \hat{X}_n }_{\mathrm{cum}}
:=
\frac{\partial}{\partial t_1}\cdots\frac{\partial}{\partial t_n}
\left.\ln\ev{\prod_{i=1}^{n} \exp\Big(t_i \hat{X}_i\Big)}\right|_{t_1=\cdots=t_n=0},
\end{equation}
which distinguishes the ordering of the operators on the lhs
and therefore is applicable
even when the operators $\{\hat{X}_i; i=1,2,\cdots,n\}$
are not commutable one another.
The expansion is then extended to triaxially-deformed intrinsic states.

\subsection{Effective Hamiltonian}
\label{subsec:effective_H}

The nuclear effective Hamiltonian has translational, rotational, parity,
and time-reversal symmetries,
with the Galilean invariance and the number conservation.
We assume that the individual terms of the Hamiltonian
also have isospin symmetries except for the Coulomb force.
The Hamiltonian is composed of the kinetic energy
$\hat{K}=\sum_i \bold{p}_i^2/(2M)$,
the effective nucleonic interaction
$\hat{V}_{\mathrm{nucl}}=\sum_{i<j} \hat{v}_{ij}$,
the Coulomb interaction between protons $\hat{V}_{\mathrm{Coul}}$,
and the center-of-mass term $\hat{H}_{\mathrm{c.m.}}=\bold{P}^2 /(2AM)$
with the total momentum $\bold{P}=\sum_i \bold{p}_i$
and the mass number $A=Z+N$,
\begin{equation}\label{eq:Hamiltonian}
  \hat{H} = \hat{K} + \hat{V}_{\mathrm{nucl}} + \hat{V}_{\mathrm{Coul}}
  - \hat{H}_{\mathrm{c.m.}}.
\end{equation}
The effective nucleonic interaction for the self-consistent MF calculations
consists of the following terms~\cite{Na03,NS02}:
\begin{equation}\label{eq:nuclear_forces}
\hat{V}_{\mathrm{nucl}}
= \hat{V}^{(\mathrm{C})} + \hat{V}^{(\mathrm{LS})} + \hat{V}^{(\mathrm{TN})}
+ \hat{V}^{(\mathrm{C\rho})}\,;\quad
\hat{V}^{(\mathrm{X})}
= \sum_{i<j} \hat{v}_{ij}^{(\mathrm{X})},\quad
(\mathrm{X}=\mathrm{C, LS, TN, C\rho}),
\end{equation}
where $\hat{V}^{(\mathrm{C})}$, $\hat{V}^{(\mathrm{LS})}$
and $\hat{V}^{(\mathrm{TN})}$
are the central, LS and tensor forces.
For the individual terms of Eq.~\eqref{eq:nuclear_forces},
we consider the following forms:
\begin{equation}\label{eq:interaction}
\begin{split}
\hat{v}_{ij}^{(\mathrm{C})}
=& \sum_{n} \left(
t^{\mathrm{(SE)}}_{n} P_{\mathrm{SE}} + t^{\mathrm{(TE)}}_{n} P_{\mathrm{TE}}
+ t^{\mathrm{(SO)}}_{n} P_{\mathrm{SO}} + t^{\mathrm{(TO)}}_{n} P_{\mathrm{TO}}
\right) f^{\mathrm{(C)}}_{n}(r_{ij}), \\
\hat{v}_{ij}^{(\mathrm{LS})}
=& \sum_{n} \left(
t^{\mathrm{(LSE)}}_{n} P_{\mathrm{TE}} + t^{\mathrm{(LSO)}}_{n} P_{\mathrm{TO}}
\right) f^{\mathrm{(LS)}}_{n}(r_{ij})\,
\bold{L}_{ij} \cdot (\bold{s}_i + \bold{s}_j), \\
\hat{v}_{ij}^{(\mathrm{TN})}
=& \sum_{n} \left(
t^{\mathrm{(TNE)}}_{n} P_{\mathrm{TE}} + t^{\mathrm{(TNO)}}_{n} P_{\mathrm{TO}}
\right) f^{\mathrm{(TN)}}_{n}(r_{ij})\,r^2_{ij} \, S_{ij}, \\
\hat{v}_{ij}^{(\mathrm{C\rho})}
=& \left(
t^{\mathrm{(SE)}}_{\rho} P_{\mathrm{SE}} \cdot \left[ \rho(\bold{r}_i) \right]^{\alpha^{(\mathrm{SE})}}
+ t^{\mathrm{(TE)}}_{\rho} P_{\mathrm{TE}} \cdot \left[ \rho(\bold{r}_i) \right]^{\alpha^{(\mathrm{TE})}}
\right) \delta(\bold{r}_{ij}),
\end{split}
\end{equation}
where
$ \bold{r}_{ij} := \bold{r}_{i} - \bold{r}_{j} $,
$ r_{ij} := |\bold{r}_{ij}| $,
$ \bold{\hat{r}}_{ij} := \bold{r}_{ij} / r_{ij} $,
$ \bold{p}_{ij} := (\bold{p}_{i} - \bold{p}_{j})/2 $,
$ \bold{L}_{ij} := \bold{r}_{ij} \times \bold{p}_{ij} $,
$S_{ij} := 4\big[3(\bold{s}_i \cdot \bold{\hat{r}}_{ij})
(\bold{s}_j \cdot \bold{\hat{r}}_{ij})
  - \bold{s}_i \cdot \bold{s}_j\big]$,
and $\rho(\mathbf{r})$ is the nucleon density.
The central density-dependent term is distinguished
from $\hat{V}^{(\mathrm{C})}$ and represented by $\hat{V}^{(\mathrm{C\rho})}$.
The projection operators on the singlet-even (SE), triplet-even (TE),
singlet-odd (SO) and triplet-odd (TO) two-nucleon states
are denoted by $P_\mathrm{Y}$
($\mathrm{Y}=\mathrm{SE},\mathrm{TE},\mathrm{SO},\mathrm{TO}$).

We adopt the semi-realistic interaction M3Y-P6 \cite{Na13,Na20,SNM16,MN18},
in which the Yukawa function
$f_{n}^{(\mathrm{X})}(r) =e^{-\mu_{n}^{(\mathrm{X})}r}/(\mu_{n}^{(\mathrm{X})}r)$ is used
for the radial functions, except for $\hat{v}_{ij}^{(\mathrm{C\rho})}$.
The longest-range term in $\hat{v}^{\mathrm{(C)}}_{ij}$ is fixed to be
that of the one-pion exchange potential (OPEP).
This central OPEP, denoted by $\hat{V}^{\mathrm{(OPEP)}}$,
is an example of spin-dependent interactions.
The values of the parameters for M3Y-P6 are given in Ref. \cite{Na13}.

\subsection{MF calculations}\label{subsec:MF_calculations}

The self-consistent MF calculations have been implemented
via the Gaussian expansion method (GEM)~\cite{Na13,Na20,SNM16,MN18,Na08}.
The generalized Bogoliubov transformation in the HFB theory
is given as~\cite{RS80,Na06}
\begin{equation}\label{eq:transUV}
\alpha^{\dagger}_i := \sum_k
\left( c^{\dagger}_k \mathsf{U}_{ki} + c_k \mathsf{V}_{ki} \right).
\end{equation}
We assume the axial, time-reversal, and parity symmetry
on the MF state $\ket{\Phi_0}$,
and then the variational parameters $\mathsf{U}_{ki}$ and $\mathsf{V}_{ki}$
are taken to be real numbers.
Adding the terms constraining the particle numbers,
we modify the Hamiltonian as
\begin{equation}\label{eq:H'}
\hat{H}' := \hat{H}-
\mu_p (\hat{N}_p - Z) - \mu_n (\hat{N}_n - N),
\end{equation}
where $\mu_p$ ($\mu_n$) is the chemical potential for protons (neutrons),
and $\hat{N}_p$ ($\hat{N}_n$) is the proton (neutron) number operator.
The energy $\ev*{\hat{H}'}{\Phi_0}$ is minimized in the HFB calculations.
We do not consider the proton-neutron pairing in this paper.

The HFB handles the pairing effects self-consistently
by taking into account the influences of the pairing
on the particle-hole channel.
However, in the HFB,
the pairing may influence the particle-hole channel
and alter the HF configuration.
For clarifying the effects of the pairing,
the HF+BCS (Bardeen-Cooper-Schrieffer) method is useful as well,
in which the HF configuration is fixed.
For analyzing the pairing effects on the pure rotational energy,
we introduce a parameter $g$ as
\begin{equation}\label{eq:H_with_g}
\ev*{\hat{H}_g} = \ev*{\hat{H}_{\mathrm{dns}}} + g\ev*{\hat{H}_{\mathrm{pair}}},
\end{equation}
where $\ev*{\hat{H}_{\mathrm{dns}}}$ consists of the terms
including only the density matrix,
while $\ev*{\hat{H}_{\mathrm{pair}}}$ is the pair energy
containing the pairing tensor~\cite{RS80}.
In the HF+BCS calculations,
the axial-HF solution has been solved self-consistently,
and the BCS equation is solved for $\ev*{\hat{H}_g}$
on top of the HF single-particle (s.p.) states.
We denote its solution by $\ket{\Phi_0}_g$.
The state $\ket{\Phi_0}_{g=0}$ corresponds to the HF state,
and the state $\ket{\Phi_0}_{g=1}$ does to the HF+BCS state
with the original Hamiltonian $\hat{H}'$.
Throughout this paper,
the parameter $g$ is employed only in the HF+BCS scheme.
We always apply $\hat{H}'$ of Eq.~\eqref{eq:H'} to the HFB calculation,
not using $\ev*{\hat{H}_g}$.

\subsection{Implementation of AMP}

In this work, the PAV has been applied for the AMP calculations
of Eq.~\eqref{eq:SJ}.
The number projection is not applied.
In reality, the intrinsic state could gradually change with increasing $J$,
often accompanied by a breakdown of the axial and the time-reversal symmetry.
While these effects can be handled in the cranking model~\cite{RS80,PT62,Ka68}
and in the VAP approaches~\cite{RS80},
they are ignored in the present study,
and we focus on the pure rotational energy,
\textit{i.e.}, the rotational energy arising from a fixed intrinsic state,
as stated in Introduction.

The overlap function $\ev*{e^{-i\hat{J}_y\beta}}{\Phi_0}$ has been calculated
from the Onishi formula~\cite{RS80,OY66,BB69,AN22}.
Concerning the sign problem of the Onishi formula,
solutions have been proposed~\cite{NW83,Ro09}
and the non-negativity of the overlap function has been proven for the HF states
under the time-reversal symmetry in Appendix~C of Ref.~\cite{AN22}.
We have here confirmed via the continuity with respect to $\beta$
that the sign of the overlap function is positive
in all the cases under consideration.

There is a problem in the density-dependent coefficients
in $\hat{v}_{ij}^{\mathrm{(C\rho)}}$
in the AMP calculations~\cite{BH08,BB21,SDRRY21,AN22}.
In the present calculations,
the standard treatment in Refs.~\cite{RER02,AN22} has been adopted,
replacing the density $\rho(\mathbf{r})$ in Eq.~\eqref{eq:interaction}
with the ``{\it generalized density}\,'' $\bar{\rho}(\mathbf{r};\beta)$,
\begin{equation}\label{eq:generalized_dens}
\bar{\rho}(\mathbf{r};\beta)
:= \sum_{\tau} \sum_{\sigma}
\frac{\ev*{\hat{\rho}(\mathbf{r}\sigma\tau)\,e^{-i \hat{J}_y \beta}}{\Phi_0}}
{\ev*{e^{-i \hat{J}_y \beta}}{\Phi_0}};\quad
\hat{\rho}(\mathbf{r}\sigma\tau)
:=\psi^\dagger(\mathbf{r}\sigma\tau) \psi(\mathbf{r}\sigma\tau),
\end{equation}
where $\psi^\dagger(\mathbf{r}\sigma \tau)$
and $\psi(\mathbf{r}\sigma \tau)$
stand for the creation and annihilation operators
for the spinor field~\cite{NO95},
which satisfy the fermionic anticommutation relations:
\begin{equation}
  \{\psi(x_1), \psi^\dagger(x_2)\} = \delta(x_1-x_2),\quad
  \{\psi(x_1), \psi(x_2)\} = \{\psi^\dagger(x_1), \psi^\dagger(x_2)\} = 0.
\end{equation}
with the shorthand notation $x_i=(\mathbf{r}_i \sigma_i \tau_i)$ ($i=1,2$)
and $\delta(x_1-x_2) := \delta(\mathbf{r}_1-\mathbf{r}_2)
\delta_{\sigma_1 \sigma_2} \delta_{\tau_1 \tau_2}$.
Although $\bar{\rho}(\mathbf{r};\beta)$ in Eq.~\eqref{eq:generalized_dens}
is a real number, owing to the time-reversal symmetry,
$\bar{\rho}^{\,\alpha}(\mathbf{r};\beta)$ is multivalued
unless the power $\alpha$ is an integer.
In the M3Y-P6 interaction,
$\alpha^{\mathrm{(SE)}}=1$ and $\alpha^{\mathrm{(TE)}}=1/3$~\cite{Na13}.
The phase of $\bar{\rho}^{\,\alpha^{\mathrm{(TE)}}}(\mathbf{r};\beta)$
has been chosen to be negative
when $\bar{\rho}(\mathbf{r};\beta)$ is negative~\cite{AN22}.

\subsection{Degree of proximity between nucleons}
\label{subsec:DoP}

To investigate the relevance of the spatial correlations between nucleons
to the rotational energy,
we consider an operator comprised of the two-body delta function,
\begin{equation}\label{eq:def_2body_delta}
\hat{D}:=\sum_{i<j} \delta(\mathbf{r}_{ij}).
\end{equation}
The expectation value $\ev*{\hat{D}}$ measures the degree
how frequently two constituent nucleons sit at an equal position,
representing the degree that two nucleons get spatially close to each other.
We call $\ev*{\hat{D}}$ \textit{degree of proximity (DoP)} in this paper.

The pair-distribution function has been employed
to investigate the spatial correlation between two particles~\cite{GV05},
such as the di-neutron correlation~\cite{MKY05}.
The pair-distribution function is defined as
\begin{equation}\label{eq:def_pair_distr}
\mathcal{G}(x_1,x_2)
:=\,\frac{\ev*{ \psi^\dagger(x_1) \psi^\dagger(x_2) \psi(x_2) \psi(x_1)}}
{\ev*{\hat{\rho}(x_1)} \ev*{\hat{\rho}(x_2)}},
\end{equation}
with $x_i=(\mathbf{r}_i \sigma_i \tau_i)$.
Notice $\mathcal{G}(x_1,x_2) = 0$ for $x_1=x_2$,
and $\mathcal{G}(x_1,x_2) \ge 0$.
The DoP is regarded as a summation of the pair-distribution function
at the same position, \textit{i.e.},
\begin{equation}\label{eq:2body_delta_approx}
\begin{split}
\ev*{\hat{D}}
&= \frac{1}{2} \sum_{\tau_1\tau_2} \sum_{\sigma_1\sigma_2} \int d^{\,3}r
\ev*{\psi^\dagger(\mathbf{r}\sigma_1\tau_1) \psi^\dagger(\mathbf{r}\sigma_2\tau_2)
  \psi(\mathbf{r}\sigma_2\tau_2) \psi(\mathbf{r}\sigma_1\tau_1)}\\
&= \frac{1}{2} \sum_{\tau_1\tau_2} \sum_{\sigma_1\sigma_2} \int d^{\,3}r
\ev*{\hat{\rho}(\mathbf{r}\sigma_1\tau_1)} \ev*{\hat{\rho}(\mathbf{r}\sigma_2\tau_2)}
\mathcal{G}(\mathbf{r}\sigma_1\tau_1,\mathbf{r}\sigma_2\tau_2).
\end{split}
\end{equation}

Since $\hat{D}$ is a rotational scalar,
the DoP for angular-momentum eigenstates $\ev*{\hat{D}}{J}$
is calculable via  Eq.~\eqref{eq:SJ}.
By applying $\hat{D} P_\mathrm{Y}$ ($\mathrm{Y}=\mathrm{SE},\mathrm{TE}$)
instead of $\hat{D}$ itself,
the DoP can be separated between the SE and TE channels.
Restricting $i$ and $j$ on the RHS of Eq.~\eqref{eq:def_2body_delta},
we can calculate the DoP for individual isospin components.

\section{Results}
\label{sec:Numerical_results}

In the present work,
the AMP of Eq.~\eqref{eq:SJ} is implemented on top of the axial MF solutions
for deformed $_{12}$Mg~\cite{Na20,SNM16} and $_{40}$Zr~\cite{MN18} nuclei,
including stable and unstable ones.
It has been established,
\textit{e.g.}, from the ratios of excitation energies
$E_{\mathrm{x}}(4^+)/E_{\mathrm{x}}(2^+)$~\cite{NNDC,DS13,SY11,PC17},
that $^{24}_{12}$Mg, $^{34-38}_{~~~~12}$Mg~ \cite{DS13},
$^{80}_{40}$Zr~\cite{LC87} and $^{100-110}_{~~~~~~\,40}$Zr~\cite{NNDC,SY11,PC17}
are well-deformed.
$^{40}_{12}$Mg lies near the neutron dripline~\cite{BA07},
and a deformed halo structure has been predicted~\cite{Na08,NT18}.
In the following,
we shall show the AMP results on top of the HFB solutions
in $^{24,34,40}_{~~~~~\,\,12}$Mg and $^{80,100,104}_{~~~~~~~~40}$Zr,
whereas we limit those on top of the HF+BCS results with varying $g$
to $^{24}_{12}$Mg and $^{80,100}_{~~~~40}$Zr,
which are suffcient for the present discussion.

\subsection{Influence of pairing on deformation}
\label{subsec:deformation}

\begin{figure}[]
\centering
\includegraphics[width=14cm,clip]{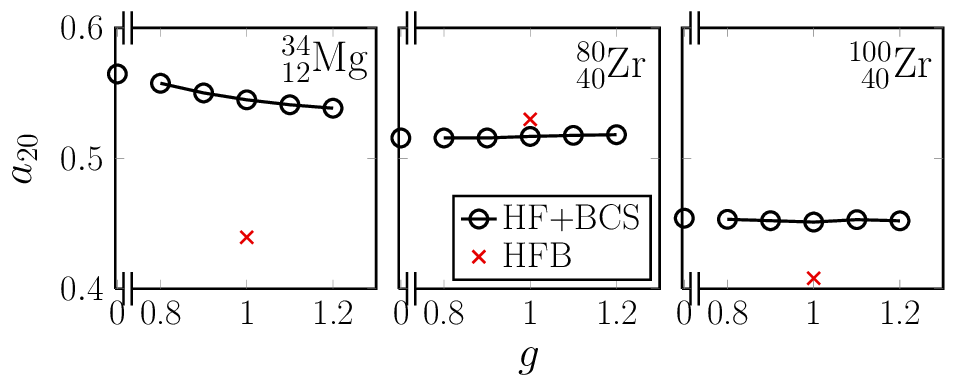}
\caption[]
{
The $g$ dependence of the deformation parameter $a_{20}$
in the HF+BCS results with $\ev*{\hat{H}_g}$
for $^{34}_{12}$Mg and $^{80,100}_{~~~~40}$Zr.
The red crosses represent the $a_{20}$ values in the HFB solutions.
}
\label{def_para_HF+BCS}
\end{figure}

We define the quadrupole deformation parameter $a_{20}$ as follows~\cite{BM98},
\begin{equation}\label{eq:def_para}
a_{20}:=\frac{q_0}{1.09 A^{5/3}\,\mathrm{fm}^2},
\end{equation}
where $q_0$ is the mass quadrupole moment of the MF state~\cite{SNM16}.
In Fig.~ \ref{def_para_HF+BCS},
the dependence of the deformation parameter $a_{20}$ on the pairing strength $g$
in the HF+BCS solutions [see Eq.~\eqref{eq:H_with_g}]
is depicted for the $^{34}_{12}$Mg and $^{80,100}_{~~~~40}$Zr nuclei.
The state $\ket{\Phi_0}_{g=0}$ corresponds with the HF state.
As $g$ grows, the pair correlations arise at the critical values,
corresponding with the normal-to-superfluid phase transition.
The critical $g$ values for proton and neutron pairs are close,
though not equal.
While $a_{20}$ is insensitive to $g$ for the $^{80,100}_{~~~~40}$Zr nuclei,
$a_{20}$ slightly decreases as $g$ grows for the $^{34}_{12}$Mg nucleus.
The values of $a_{20}$ for the HFB minima are also shown.
The $a_{20}$ values for the HF+BCS solutions are close to
that of the HFB solution for the $^{80}_{40}$Zr nucleus,
while they do not match well for the $^{34}_{12}$Mg and $^{100}_{~40}$Zr nuclei,
indicating influence of the pairing on the HF configurations. 
Whereas the $a_{20}$ values for the HFB solutions are smaller
than those for the HF solutions in $^{34}_{12}$Mg and $^{100}_{~40}$Zr,
the opposite is realized at $^{80}_{40}$Zr.
This enhancement of deformation at $^{80}_{40}$Zr
takes place owing to the s.p. levels near the Fermi energy.
The pairing moves a portion of neutrons
at the highest occupied level with $\Omega^\pi=5/2^+$,
which originates from the $0g_{9/2}$ spherical orbit,
to the lowest unoccupied level with $\Omega^\pi=1/2^+$,
by which the prolate deformation is slightly enhanced.
Table~\ref{tab:def_para} presents the $a_{20}$ values
for the HF and HFB solutions
at their lowest minima for $^{24}_{12}$Mg, $^{40}_{12}$Mg and $^{104}_{~40}$Zr.

\begin{table}
\begin{center}
  \caption{$a_{20}$ for the HF and HFB solutions at their lowest minima
    for $^{24}_{12}$Mg, $^{40}_{12}$Mg and $^{104}_{~40}$Zr.
\label{tab:def_para}}
\begin{tabular}{crr}
\hline\hline
nuclide & HF & HFB \\
\hline
$^{24}_{12}$Mg & $0.54$ & $0.54$ \\
$^{40}_{12}$Mg & $0.47$ & $0.43$ \\
$^{104}_{~40}$Zr & $0.46$ & $0.43$ \\
\hline\hline
\end{tabular}
\end{center}
\end{table}

\subsection{Comparison of $E_{\mathrm{x}}(2^+)$ with rigid-rotor model
  and experiment}
\label{subsec:Exp_RR}

Let us denote the expectation value of Eq.~\eqref{eq:SJ}
measured from that of the g.s. by
\begin{equation}\label{eq:S_x}
\mathcal{S}_{\mathrm{x}}(J^+)
:= \ev*{\hat{\mathcal{S}}}{J} - \ev*{\hat{\mathcal{S}}}{0}.
\end{equation}
We explicitly attach the parity quantum number ($+$) on the LHS.
For $\mathcal{\hat{S}}=\hat{H}'$,
$\mathcal{S}_{\mathrm{x}}(J^+)$ corresponds with the excitation energy
(\textit{i.e.}, the pure rotational energy),
\begin{equation}\label{eq:E_x}
E_{\mathrm{x}}(J^+)
= \ev*{\hat{H}'}{J} - \ev*{\hat{H}'}{0}.
\end{equation}

\begin{figure}[]
\centering
\includegraphics[width=15cm,clip]{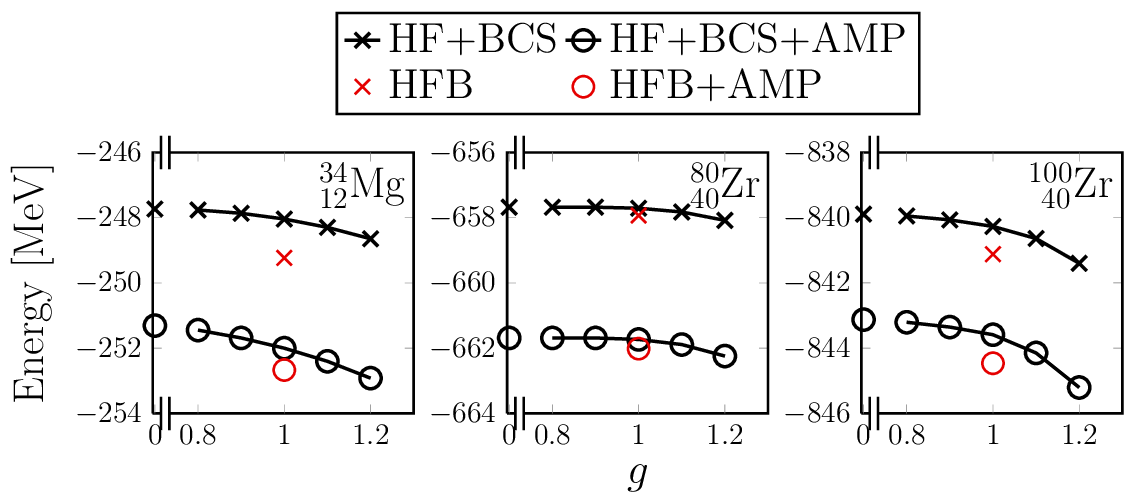}
\caption[]
{
  The unprojected and the projected energies
  for $^{34}_{12}$Mg and $^{80,100}_{~~~~40}$Zr.
  The black crosses represent the unprojected HF+BCS energies
  depending on $g$,
  and the red circles do the corresponding projected energies $E(0^+)$.
  The red crosses (circles) represent the unprojected (projected) HFB energies.
}
\label{energy_HF+BCS}
\end{figure}

In Fig.~\ref{energy_HF+BCS},
the $g$ dependence of the unprojected
and the projected ($E(0^+)=\ev*{\hat{H}'}{0}$) g.s. energies is shown
for the HF+BCS solutions of $^{34}_{12}$Mg and $^{80,100}_{~~~~40}$Zr.
Up to the critical $g$ value where the pair correlations arise,
the energies are equal to the HF case.
As $g$ increases from the critical values,
both the unprojected and projected energies decrease for the HF+BCS solutions.
The unprojected and projected energies for the HFB minima are also shown.
Irrespective of the unprojected or the projected energies,
the energies for the HFB solutions are close to those for the HF+BCS ones
in the region between $g=1.1$ and $1.2$.

\begin{figure}[]
\centering
\includegraphics[width=14cm,clip]{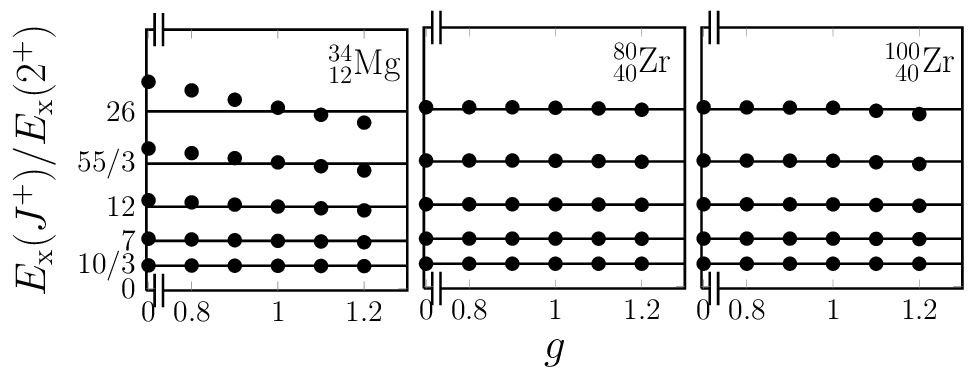}
\caption[]
{
  $E_{\mathrm{x}}(J^+)/E_{\mathrm{x}}(2^+)$
  for the HF+BCS solutions of $^{34}_{12}$Mg and $^{80,100}_{~~~~40}$Zr.
  The lines display the rigid-rotor values $J(J+1)/6$.
}
\label{ratio_HF+BCS}
\end{figure}

As it is equal to $J(J+1)/6$ in the rigid-rotor model,
the ratio $E_{\mathrm{x}}(J^+)/E_{\mathrm{x}}(2^+)$ can be a measure
of how well the rotational band develops.
The $g$ dependence of $E_{\mathrm{x}}(J^+)/E_{\mathrm{x}}(2^+)$
is shown for the $^{34}_{12}$Mg and $^{80,100}_{~~~~40}$Zr nuclei
in Fig.~\ref{ratio_HF+BCS}.
The ratios $E_{\mathrm{x}}(J^+)/E_{\mathrm{x}}(2^+)$
are insensitive to $g$ and close to the $J(J+1)/6$ lines
except for high $J\,(\gtrsim 10)$ at $^{34}_{12}$Mg.
For the $^{34}_{12}$Mg nucleus,
the ratios $E_{\mathrm{x}}(J^+)/E_{\mathrm{x}}(2^+)$ decrease for high $J$
as $g$ increases.
The deviation from the $J(J+1)/6$ lines indicates
that the higher-$c_{2n}$ terms are not negligible in Eq.~\eqref{eq:JSJ_Lambda}.

\begin{figure}[]
\centering
\includegraphics[width=15cm,clip]{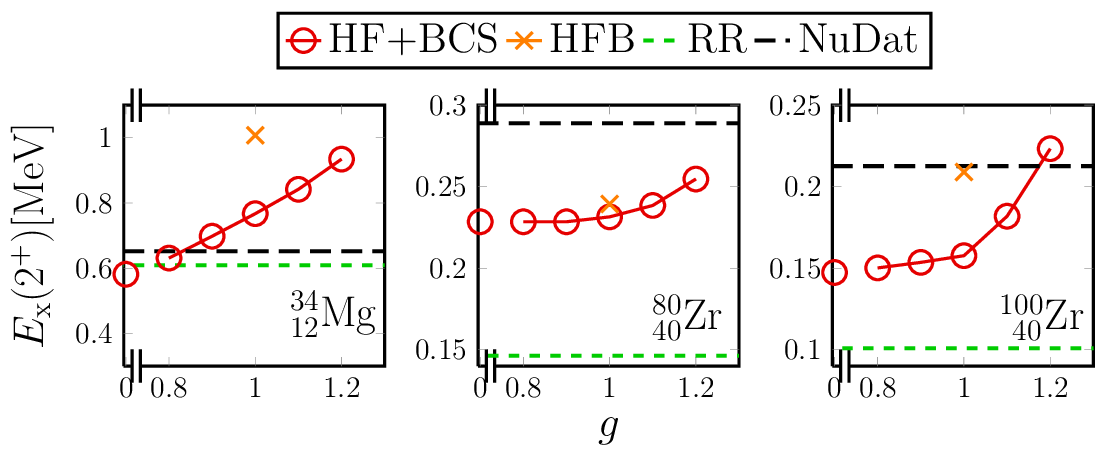}
\caption[]
{
  $E_{\mathrm{x}}(2^+)$ for the HF+BCS solutions
  of $^{34}_{12}$Mg and $^{80,100}_{~~~~40}$Zr,
  which is represented by the red circles.
  The orange crosses represent the values for the HFB solutions.
  The black dashed lines display the experimental values~\cite{NNDC},
  and the green dashed lines are the rigid-rotor values
  in Eq.~\eqref{eq:RR_approx}~\cite{BM98}.
}
\label{exc_energy_HF+BCS_HFB}
\end{figure}

Figure~\ref{exc_energy_HF+BCS_HFB} shows
the $g$ dependence of the excitation energies $E_{\mathrm{x}}(2^+)$
for the HF+BCS solutions of $^{34}_{12}$Mg and $^{80,100}_{~~~~40}$Zr.
Those for the HFB solutions are also shown.
As $g$ increases, $E_{\mathrm{x}}(2^+)$ gets higher;
the pair correlations reduce the MoI.
The excitation energies $E_{\mathrm{x}}(2^+)$ of the HF+BCS solutions
between $g=1.0$ and $1.2$
are close to those of the HFB solutions.
The $E_{\mathrm{x}}(2^+)$ values for the HFB solutions in these nuclei
are about 1.5\,--\,2 times higher than the rigid-rotor value~\cite{BM98},
\begin{equation}\label{eq:RR_approx}
E_{\mathrm{x}}^{(\mathrm{RR})}(J^+)
=\,\frac{J(J+1)}{2 \, \mathcal{I}^{\mathrm{(RR)}}};\quad
\mathcal{I}^{\mathrm{(RR)}}
\approx\,0.0138 \, A^{5/3} [\mathrm{MeV}^{-1}].
\end{equation}
Compared to the experimental values,
the $E_{\mathrm{x}}(2^+)$ value is high
in the HFB solution of the $^{34}_{12}$Mg nucleus,
while slightly low in $^{80,100}_{~~~~40}$Zr.

\subsection{Influence of higher-order terms in cumulant expansion}
\label{subsec:higher-order}

\begin{figure}[]
\centering
\includegraphics[width=14cm,clip]{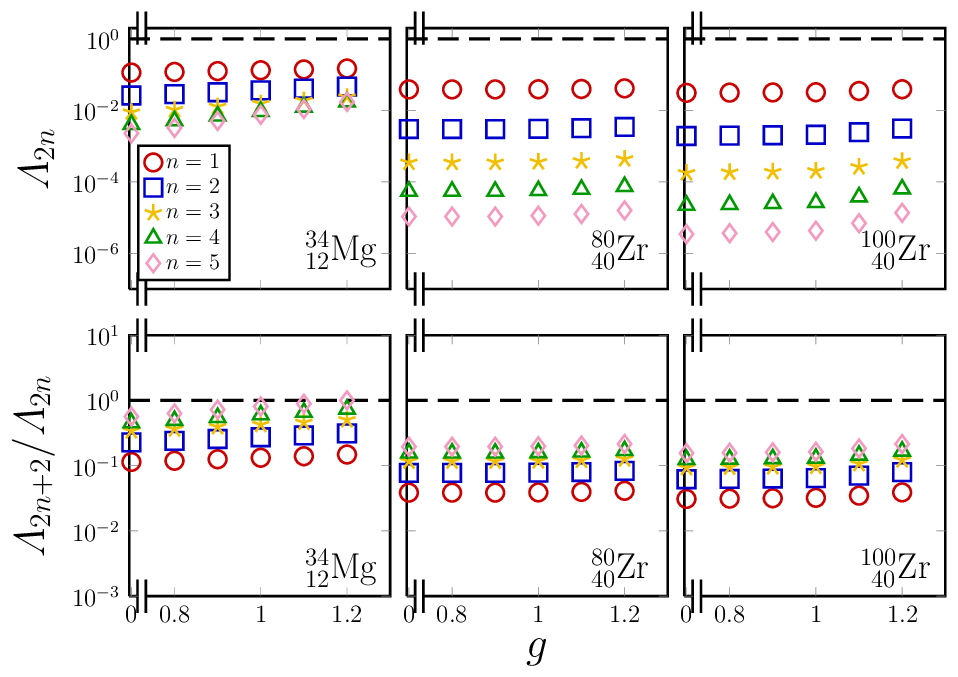}
\caption[]
{
  $\varLambda_{2n}$ and $\varLambda_{2n+2}/\varLambda_{2n}$
  in Eq.~\eqref{eq:N_2n_Lambda_2n}
  for the HF+BCS results of $^{34}_{12}$Mg and $^{80,100}_{~~~~40}$Zr.
  The symbols correspond with the $n$ values indicated in the inset.
}
\label{Lambda_2n_HF+BCS}
\end{figure}

We next investigate pairing effects on higher-order terms
in Eq.~\eqref{eq:JSJ_Lambda}.
In Fig.~\ref{Lambda_2n_HF+BCS},
the $g$ dependence of $\varLambda_{2n}$ and $\varLambda_{2n+2}/\varLambda_{2n}$
in Eq.~\eqref{eq:N_2n_Lambda_2n} is shown
for the HF+BCS solutions of $^{34}_{12}$Mg and $^{80,100}_{~~~~40}$Zr.
If $\varLambda_{2n+2}/\varLambda_{2n}$ stays small,
both the $J(J+1)$ rule and the PY approximation are validated.
For growing $g$,
$\varLambda_{2n}$ and $\varLambda_{2n+2}/\varLambda_{2n}$ slightly increase
for fixed $n$.

\begin{figure}[]
\centering
\includegraphics[width=14cm,clip]{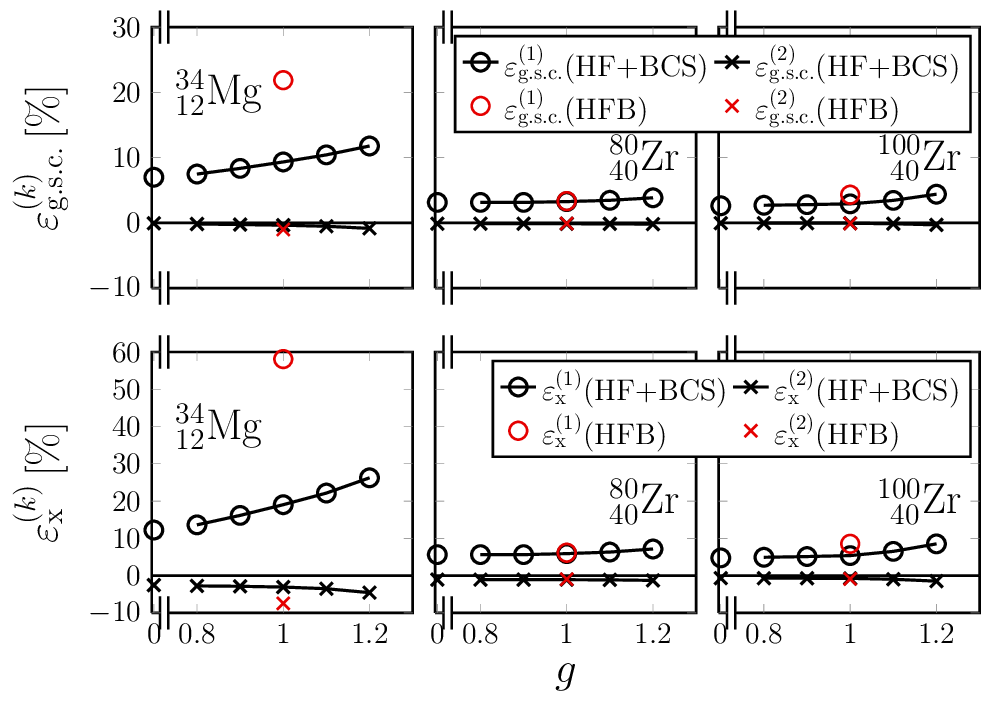}
\caption[]
   {$\varepsilon^{(k)}_{\mathrm{g.s.c.}}$ and $\varepsilon^{(k)}_{\mathrm{x}}$
     ($k=1,2$) in the HF+BCS and HFB solutions
     of $^{34}_{12}$Mg and $^{80,100}_{~~~~40}$Zr.}
\label{PYmoi_HF+BCS}
\end{figure}

The g.s. correlation is defined and expanded by $s_{2n}$
for $\hat{\mathcal{S}}=\hat{H}'$ as
\begin{equation}\label{eq:def_g.s.c.}
\varDelta E_{\mathrm{g.s.c.}} := \ev*{\hat{H}'}{\Phi_0} - \ev*{\hat{H}'}{0}
=\displaystyle -\sum_{n=1}^{\infty} s_{2n} \varLambda_{2n}.
\end{equation}
To view the influence of the higher-order terms,
we have calculated the following quantities:
\begin{subequations}\label{eq:relative_error}
\begin{align}
&\varepsilon^{(k)}_{\mathrm{g.s.c.}}
  :=\,\frac{\displaystyle -\sum_{n=1}^{k} s_{2n} \varLambda_{2n}
    - \varDelta E_{\mathrm{g.s.c.}}}
  {\varDelta E_{\mathrm{g.s.c.}}},
\label{eq:relative_error_g.s.c.}\\
&\varepsilon^{(k)}_{\mathrm{x}}
:=\,\frac{\displaystyle 3 \sum_{n=1}^{k} s_{2n}
  \left[- \frac{1}{2}(\varLambda_{2n+2} - \varLambda_{2n} \varLambda_{2}) \right]
  - E_{\mathrm{x}}(2^+)}
      {E_{\mathrm{x}}(2^+)}.
\label{eq:relative_error_Ex}
\end{align}
\end{subequations}
We calculate $s_2$ and $s_4$ from Eq.~\eqref{eq:def_S01_s2n}
via numerical differentiation.
The values of $\varepsilon^{(k)}_{\mathrm{g.s.c.}}$
and $\varepsilon^{(k)}_{\mathrm{x}}$ ($k=1,2$) are shown
for $^{34}_{12}$Mg and $^{80,100}_{~~~~40}$Zr in Fig.~\ref{PYmoi_HF+BCS}.
Insensitive to $g$,
$\varepsilon^{(1)}_{\mathrm{g.s.c.}}$ and $\varepsilon^{(1)}_{\mathrm{x}}$
almost vanish for $^{80,100}_{~~~~40}$Zr.
Namely, the contributions of the higher-$s_{2n}$ terms
to $\varDelta E_{\mathrm{g.s.c.}}$ and $E_{\mathrm{x}}(2^+)$ are negligible,
and the PY formula (Eq.~\eqref{eq:def_moi} truncated at $n=1$) is good.
On the other hand,
the $\varepsilon^{(1)}_{\mathrm{g.s.c.}}$
and $\varepsilon^{(1)}_{\mathrm{x}}$ values become larger
as $g$ increases in the $^{34}_{12}$Mg nucleus.
The pair correlations enhance
the higher-$s_{2n}$ terms of the cumulant expansion,
the terms including $s_4$ in practice,
in the g.s. correlations and the MoI
of this light nucleus.
The $\varepsilon^{(2)}_{\mathrm{g.s.c.}}$
and $\varepsilon^{(2)}_{\mathrm{x}}$ values are vanishing.

In Fig.~\ref{PYmoi_HF+BCS},
the values of $\varepsilon^{(k)}_{\mathrm{g.s.c.}}$
and $\varepsilon^{(k)}_{\mathrm{x}}$ for $k=1,2$ are also shown
for the HFB solutions of $^{34}_{12}$Mg and $^{80,100}_{~~~~40}$Zr.
The results are similar to the HF+BCS cases.
While the $\varepsilon^{(1)}_{\mathrm{g.s.c.}}$
and $\varepsilon^{(1)}_{\mathrm{x}}$ values are less than 10\%
for $^{80,100}_{~~~~40}$Zr,
the $s_4$ terms are sizable for the $^{34}_{12}$Mg nucleus.
This consequence is qualitatively similar
also to the HF case in Ref.~\cite{AN22}.

\subsection{Contribution of constituent terms of effective Hamiltonian}
\label{subsec:contribution}

In Ref.~\cite{AN22},
we analyzed the composition of the pure rotational energy
of the axial-HF solutions.
In this subsection, we present influences of the pairing
on the composition of the pure rotational energy.
By taking $\hat{\mathcal{S}}$ to be constituent terms of $\hat{H}'$,
the $\mathcal{S}_{\mathrm{x}}(J^+)$ values of Eq.~\eqref{eq:S_x}
yield their contributions to the rotational energy.
In the following, $\hat{\mathcal{S}}$ is an element of the following set,
\begin{equation}\label{eq:S_in}
  \hat{\mathcal{S}} \in \{ \hat{H}',\hat{K},\hat{V}^{\mathrm{(C)}},
  \hat{V}^{\mathrm{(LS)}},\hat{V}^{\mathrm{(TN)}},\hat{V}^{\mathrm{(C\rho)}},
  \hat{V}^{\mathrm{(OPEP)}},\hat{H}_{\mathrm{pair}} \}.
\end{equation}
Each element has been defined in Sec.~\ref{sec:theory}.
Since $\ev*{\hat{N}_p}{J}\ne Z$ and $\ev*{\hat{N}_n}{J}\ne N$,
the chemical-potential terms in Eq.~\eqref{eq:H'}
have contributions to the rotational energy,
which should be attributed to other terms of the Hamiltonian
if the particle-number projection is simultaneously implemented.
However, they are insignificant,
staying within $-0.5-17\,\%$ for the MF states under investigation.

\begin{figure}[]
\centering
\includegraphics[width=14cm,clip]{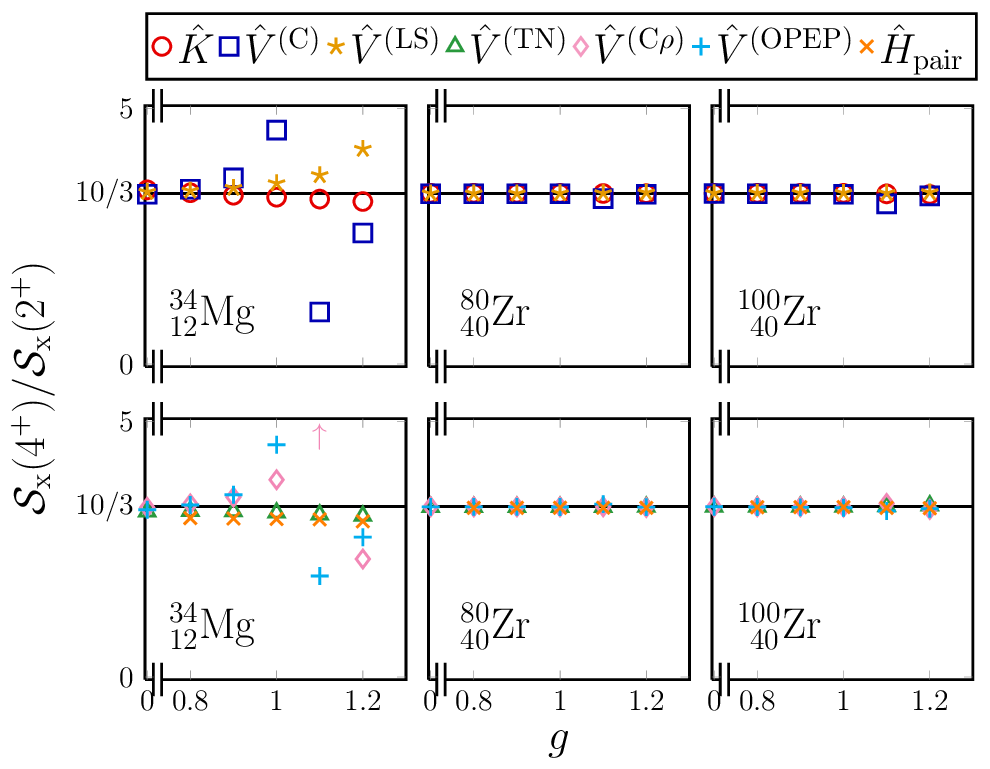}
\caption[]
{
$\mathcal{S}_{\mathrm{x}}(4^+)/\mathcal{S}_{\mathrm{x}}(2^+)$ for
$\hat{\mathcal{S}}=\hat{K}$ (red circles),
$\hat{V}^{\mathrm{(C)}}$ (blue squares),
$\hat{V}^{\mathrm{(LS)}}$ (yellow stars),
$\hat{V}^{\mathrm{(TN)}}$ (green triangles),
$\hat{V}^{\mathrm{(C\rho)}}$ (pink diamonds) and
$\hat{V}^{\mathrm{(OPEP)}}$ (sky-blue pluses)
in the HF+BCS results of $^{34}_{12}$Mg and $^{80,100}_{~~~~40}$Zr.
$\mathcal{S}_{\mathrm{x}}(4^+)/\mathcal{S}_{\mathrm{x}}(2^+)$
for $\hat{H}_{\mathrm{pair}}$
are represented by the orange crosses.
The rigid-rotor value $10/3$ is displayed by the horizontal lines.
}
\label{ratio_divide_HF+BCS}
\end{figure}

\begin{figure}[]
\centering
\includegraphics[width=14cm,clip]{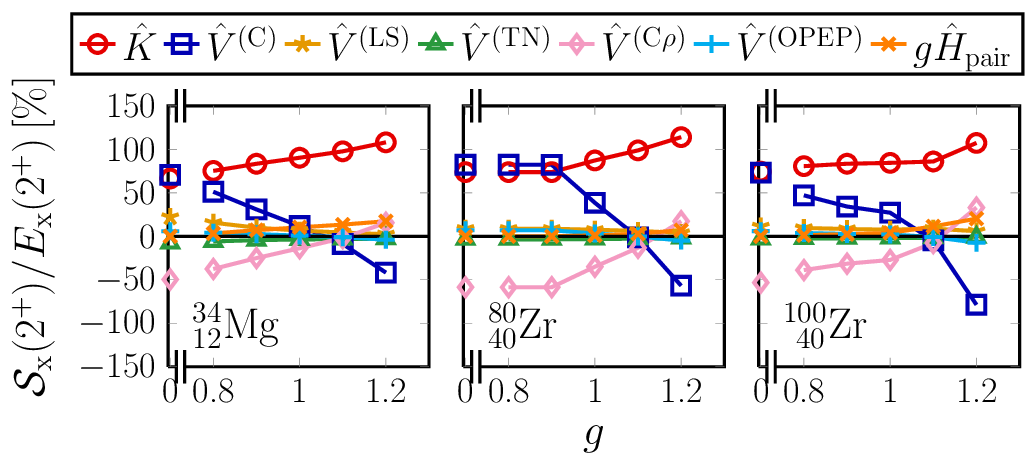}
\caption[]
{The ratios $\mathcal{S}_{\mathrm{x}}(2^+)/E_{\mathrm{x}}(2^+)$.
See Fig. \ref{ratio_divide_HF+BCS} for conventions.}
\label{percentage_HF+BCS}
\end{figure}

In Fig. \ref{ratio_divide_HF+BCS},
the $g$ dependence of $\mathcal{S}_{\mathrm{x}}(4^+)/\mathcal{S}_{\mathrm{x}}(2^+)$,
the ratios given by the constituent terms of the effective Hamiltonian,
is shown for the $^{34}_{12}$Mg and $^{80,100}_{~~~~40}$Zr nuclei.
Contributions of the pairing tensors are excluded
except for $\hat{H}_{\mathrm{pair}}$.
The ratios are close to $10/3$ for $^{80,100}_{~~~~40}$Zr,
almost independent of $g$ and $\hat{\mathcal{S}}$.
In $^{34}_{12}$Mg,
the ratios $\mathcal{S}_{\mathrm{x}}(4^+)/\mathcal{S}_{\mathrm{x}}(2^+)$
for $\hat{\mathcal{S}}=\hat{K}$, $\hat{V}^{\mathrm{(TN)}}$
and $\hat{H}_{\mathrm{pair}}$
are also almost independent of $g$,
while those for $\hat{V}^{\mathrm{(C)}}$, $\hat{V}^{\mathrm{(LS)}}$,
$\hat{V}^{\mathrm{(C\rho)}}$ and $\hat{V}^{\mathrm{(OPEP)}}$
become deviating from $10/3$ as $g$ increases.
The irregular behavior for $\hat{V}^{\mathrm{(C)}}$ and $\hat{V}^{\mathrm{(OPEP)}}$
observed at $g\approx 1.1$ in $^{34}_{12}$Mg
happens because $\mathcal{S}_{\mathrm{x}}(2^+)\approx 0$ in this region,
with sign inversion (see Fig.~\ref{percentage_HF+BCS}).

As long as any $\mathcal{S}_{\mathrm{x}}(J^+)/\mathcal{S}_{\mathrm{x}}(2^+)$
is close to $J(J+1)/6$,
$\mathcal{S}_{\mathrm{x}}(J^+)$ is well described
by $\mathcal{I}[\hat{\mathcal{S}}]$ in Eq.~\eqref{eq:J(J+1)},
and it is sufficient to inspect $\mathcal{S}_{\mathrm{x}}(2^+)$
in analyzing the rotational energy.
The contributions of the constituent terms of the effective Hamiltonian
to the total rotational energy
are represented by $\mathcal{S}_{\mathrm{x}}(2^+)/E_{\mathrm{x}}(2^+)$.
In Fig. \ref{percentage_HF+BCS},
the $g$ dependence of $\mathcal{S}_{\mathrm{x}}(2^+)/E_{\mathrm{x}}(2^+)$ is shown
for $^{34}_{12}$Mg and $^{80,100}_{~~~~40}$Zr.
As $g$ increases from the critical points,
the $\mathcal{S}_{\mathrm{x}}(2^+)/E_{\mathrm{x}}(2^+)$ values
for $\hat{\mathcal{S}}=\hat{K}$, $\hat{V}^{\mathrm{(C)}}$ and $\hat{V}^{\mathrm{(C\rho)}}$
vary;
decrease for $\hat{\mathcal{S}}=\hat{V}^{\mathrm{(C)}}$,
while increase for $\hat{V}^{\mathrm{(C\rho)}}$.
Even their signs are inverted near $g=1.1$.
As $g$ increases,
the contribution of $g\hat{H}_{\mathrm{pair}}$ to the rotational energy
is enhanced,
which is dominated by the central force.
The contributions of $\hat{V}_{\mathrm{Coul}}$, $\hat{H}_{\mathrm{c.m.}}$ and $\hat{V}^{\mathrm{(OPEP)}}$
to $E_{\mathrm{x}}(2^+)$ are $\pm10\%$ at most.

\begin{figure}[]
\centering
\includegraphics[width=10cm,clip]{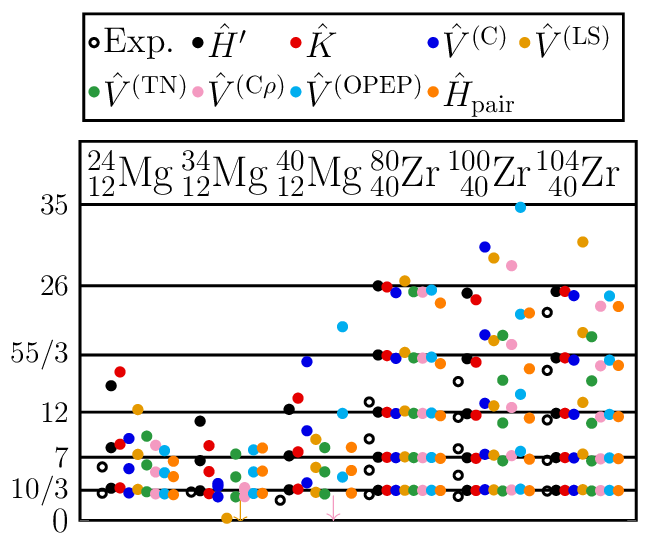}
\caption[]
{
  $\mathcal{S}_{\mathrm{x}}(J^+)/\mathcal{S}_{\mathrm{x}}(2^+)$
  for the HFB solutions of the $^{24,34,40}_{~~~~~\,\,12}$Mg
  and $^{80,100,104}_{~~~~~~~~40}$Zr nuclei.
  See Fig.~\ref{ratio_divide_HF+BCS} for conventions.
  The experimental values of $E_{\mathrm{x}}(J^+)/E_{\mathrm{x}}(2^+)$
  are taken from Refs.~\cite{NNDC,DS13,CF19}.
}
\label{ratio_HFB}
\end{figure}

Figure~\ref{ratio_HFB} shows the ratios $\mathcal{S}_{\mathrm{x}}(J^+)/\mathcal{S}_{\mathrm{x}}(2^+)$
for the HFB solutions of the $^{24,34,40}_{~~~~~\,\,12}$Mg and $^{80,100,104}_{~~~~~~~~40}$Zr nuclei,
all of which have prolate shapes.
The ratios $E_{\mathrm{x}}(4^+)/E_{\mathrm{x}}(2^+)$ obtained in the present work
are close to those of the experiments and $10/3$ except at $^{40}_{12}$Mg.
For the deformed $_{40}$Zr nuclei,
the ratios of the constituent terms $\mathcal{S}_{\mathrm{x}}(J^+)/\mathcal{S}_{\mathrm{x}}(2^+)$
are also close to $J(J+1)/6$,
although less close than in the HF case.
On the other hand,
some ratios $\mathcal{S}_{\mathrm{x}}(J^+)/\mathcal{S}_{\mathrm{x}}(2^+)$
do not obey the $J(J+1)$ rule for the $_{12}$Mg nuclei.
We find several cases in which
even $\ev*{\hat{\mathcal{S}}}{J}$ is not monotonic for $J$,
\textit{e.g.}, with $\mathcal{S}_{\mathrm{x}}(4^+)/\mathcal{S}_{\mathrm{x}}(2^+)<1$.

\begin{figure}[]
\centering
\includegraphics[width=10cm,clip]{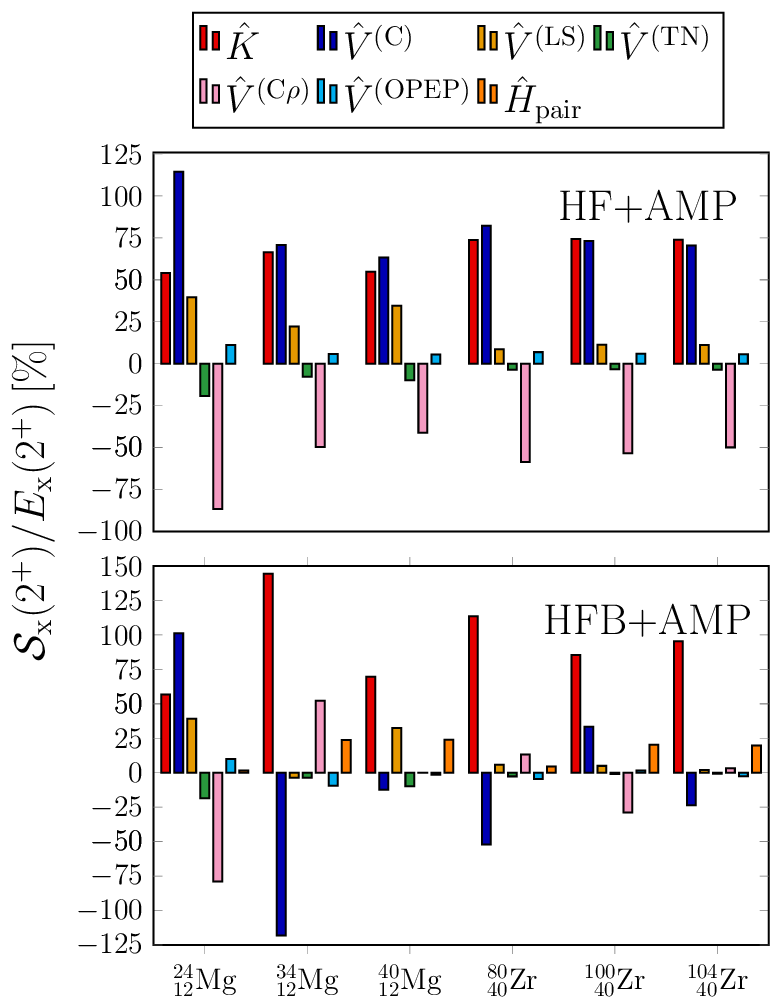}
\caption[]
{
  $\mathcal{S}_{\mathrm{x}}(2^+) / E_{\mathrm{x}}(2^+)$
  for the HFB solutions (lower panel) of the $^{24,34,40}_{~~~~~\,\,12}$Mg
  and $^{80,100,104}_{~~~~~~~~40}$Zr nuclei,
  in comparison with those for the HF solutions (upper panel)~\cite{AN22}.
}
\label{percentage_HFB_pro}
\end{figure}

In Fig.~\ref{percentage_HFB_pro},
$\mathcal{S}_{\mathrm{x}}(2^+)/E_{\mathrm{x}}(2^+)$ are shown
for the HFB solutions of the $^{24,34,40}_{~~~~~\,\,12}$Mg
and $^{80,100,104}_{~~~~~~~~40}$Zr nuclei.
The results significantly depend on the nuclei.
The contribution of the kinetic energy $\hat{\mathcal{S}}=\hat{K}$
is large.
It is remarked that those of the interactions $\hat{V}^{\mathrm{(C)}}$ and $\hat{V}^{\mathrm{(C\rho)}}$
are scattered,
in sharp contrast to the results for the deformed HF solutions
in Fig.~3 of Ref.~\cite{AN22}.
For the HF solutions,
the composition of the pure rotational energy for the well-deformed heavy nuclei
is insensitive to nuclides and deformation.
However, as elucidated in Fig. \ref{percentage_HF+BCS},
$\mathcal{S}_{\mathrm{x}}(2^+)/E_{\mathrm{x}}(2^+)$ is sensitive to the pairing.
Since the degree of the pair correlations depends on nuclei,
the components of the pure rotational energy of nuclei also do,
even for the well-deformed heavy nuclei.

\subsection{Angle dependence of integrands of Eq.~\protect\eqref{eq:SJ}}
\label{subsec:angle_dep}

In Ref.~\cite{AN22},
we showed that the overlap functions
$\ev*{e^{-i\hat{J}_y \beta}}{\Phi_0}$ and $\mathcal{S}^{01}(\beta)$
in Eq.~\eqref{eq:def_S01_s2n} are relevant
to the $J(J+1)$ rule and the composition of the rotational energy.
The dependence of the overlap functions on the angle $\beta$
will be instructive in the HF+BCS and HFB cases, as well.

\begin{figure}[]
\centering
\includegraphics[width=14cm,clip]{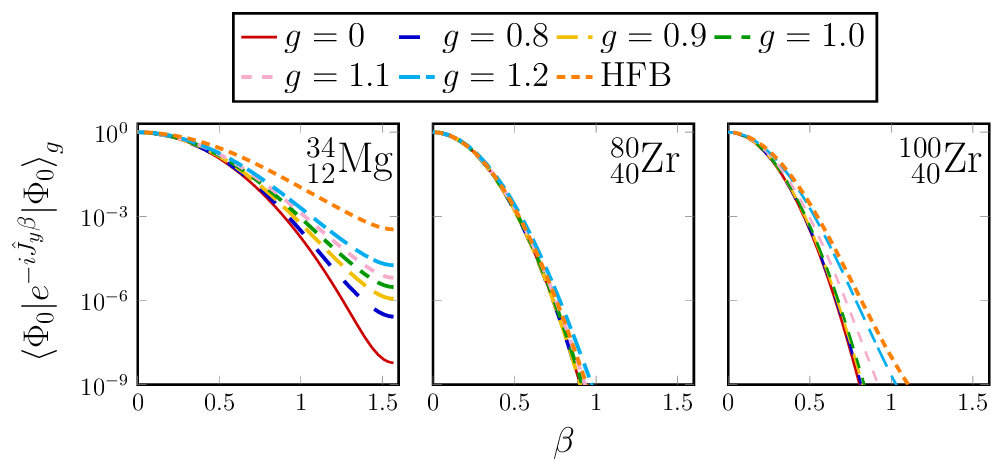}
\caption[]
{
  The overlap functions $\ev*{e^{-i\hat{J}_y\beta}}{\Phi_0}_g$
  in the HF+BCS and HFB solutions of $^{34}_{12}$Mg and $^{80,100}_{~~~~40}$Zr.
  The individual lines correspond to the $g$ values shown in the inset.
}
\label{onishi_HF+BCS}
\end{figure}

Figure~\ref{onishi_HF+BCS} shows the $g$ dependence
of the overlap functions $\ev*{e^{-i\hat{J}_y\beta}}{\Phi_0}_g$
for the HF+BCS solutions of $^{34}_{12}$Mg and $^{80,100}_{~~~~40}$Zr.
For $^{80,100}_{~~~~40}$Zr,
$\ev*{e^{-i\hat{J}_y\beta}}{\Phi_0}_g$ have sharp peaks near $\beta=0$,
and become slightly broader as $g$ increases.
In contrast, $\ev*{e^{-i\hat{J}_y\beta}}{\Phi_0}_g$ have broad peaks
near $\beta=0$
and getting much broader for increasing $g$
for the $^{34}_{12}$Mg nucleus.

\begin{figure}[]
\centering
\includegraphics[width=14cm,clip]{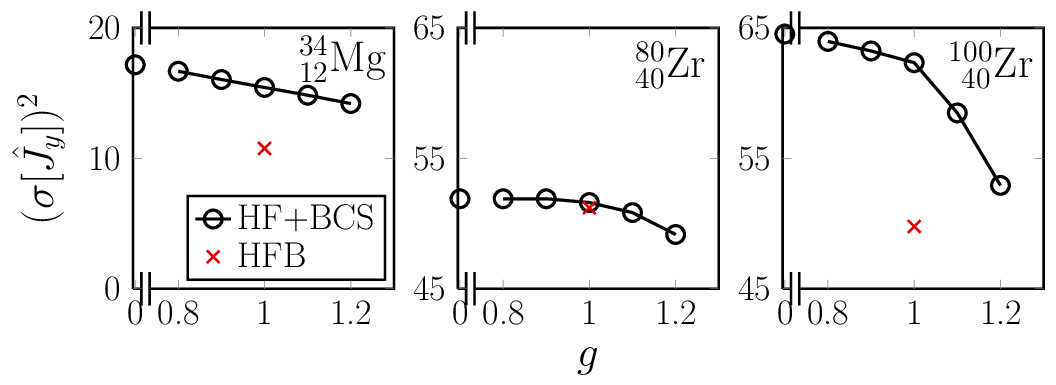}
\caption[]
  {$(\sigma[\hat{J}_y])^2$ in the HF+BCS and HFB solutions
   of $^{34}_{12}$Mg and $^{80,100}_{~~~~40}$Zr.}
\label{fluctuation_def_para_HF+BCS}
\end{figure}

Figure~\ref{onishi_HF+BCS} also depicts $\ev*{e^{-i\hat{J}_y\beta}}{\Phi_0}$
for the HFB solutions of $^{34}_{12}$Mg and $^{80,100}_{~~~~40}$Zr.
The overlap functions $\ev*{e^{-i\hat{J}_y\beta}}{\Phi_0}$
for $^{80,100}_{~~~~40}$Zr
have sharper peaks near $\beta=0$ than those for $^{34}_{12}$Mg.
The broad peak near $\beta=0$ makes
the ratios $\mathcal{S}_{\mathrm{x}}(J^+) / \mathcal{S}_{\mathrm{x}}(2^+)$
deviate from $J(J+1)/6$ in Fig.~\ref{ratio_HFB},
as analogous arguments given in Ref.~\cite{AN22}.
The curvature of $\ev*{e^{-i \hat{J}_y \beta}}{\Phi_0}$ at $\beta=0$ is
equal to the variance $(\sigma[\hat{J}_y])^2$ apart from the sign
(see Eq.~\eqref{eq:def_C_A-B}),
\begin{equation}\label{eq:sigma}
-\left. \frac{d^{\,2}}{d\beta^{\,2}}\,
\ev*{e^{-i \hat{J}_y \beta}}{\Phi_0} \, \right| _{\beta=0}
= (\sigma [ \hat{J}_y ])^2.
\end{equation}
In~Fig.~\ref{fluctuation_def_para_HF+BCS},
we show $(\sigma[\hat{J}_y])^2$ for $^{34}_{12}$Mg and $^{80,100}_{~~~~40}$Zr.
Regardless of nuclei,
the $(\sigma[\hat{J}_y])^2$ values decrease as $g$ increases,
and $(\sigma[\hat{J}_y])^2$ of the HFB solutions are smaller
than that of the HF solutions.
The $^{100}_{~40}$Zr nucleus exemplifies
that $a_{20}$ does not well correlate to $\sigma[\hat{J}_y]$ straightforwardly;
$(\sigma [ \hat{J}_y ])^2$ substantially decreases as $g$ grows,
while $a_{20}$ shown in Fig.~\ref{def_para_HF+BCS} is insensitive to $g$.

\begin{figure}[]
\centering
\includegraphics[width=14cm,clip]{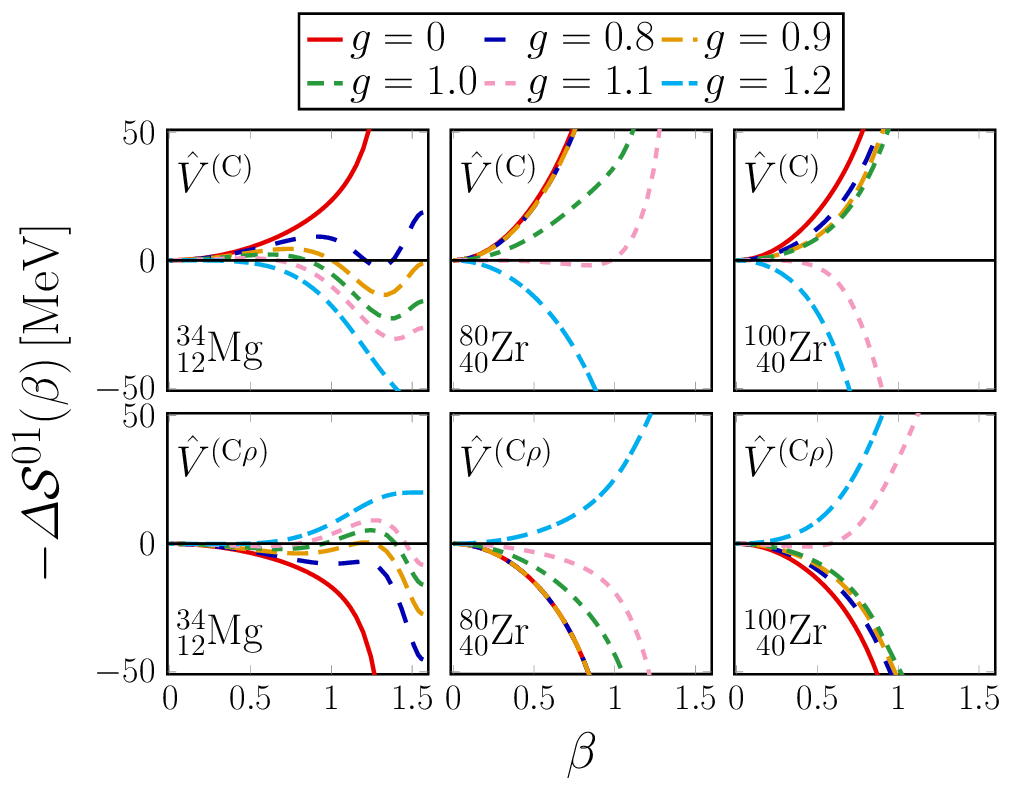}
\caption[]
  {$-\varDelta \mathcal{S}^{01}(\beta)$
  for $\hat{\mathcal{S}}=\hat{V}^{\mathrm{(C)}}$ and $\hat{V}^{\mathrm{(C\rho)}}$
  in the HF+BCS solutions of $^{34}_{12}$Mg and $^{80,100}_{~~~~40}$Zr.
  The individual lines correspond to the $g$ values shown in the inset.}
\label{beta_dep_int_HF+BCS}
\end{figure}

We next investigate the inversion near $g=1.1$
for $\hat{\mathcal{S}}=\hat{V}^{\mathrm{(C)}}$ and $\hat{V}^{\mathrm{(C\rho)}}$
observed in Fig.~\ref{percentage_HF+BCS},
at the level of the overlap functions.
By using Eq.~\eqref{eq:def_S01_s2n},
we define $\varDelta \mathcal{S}^{01}(\beta)$ as
\begin{equation}\label{eq:Delta_S01}
\varDelta \mathcal{S}^{01}(\beta)
:= \mathcal{S}^{01}(\beta)-\mathcal{S}^{01}(\beta=0).
\end{equation}
The curvature of $-\varDelta \mathcal{S}^{01}(\beta)$ is related
to the correlation function $C[ \hat{\mathcal{S}}, \hat{J}_y^{\,2} ]$,
\begin{equation}\label{eq:DeltaS01_correlation}
\left. -\, \frac{d^{\,2}}{d\beta^{\,2}}\,
\varDelta \mathcal{S}^{01}(\beta) \right| _{\beta=0}
= C[\hat{\mathcal{S}}, \hat{J}_y^{\,2}].
\end{equation}
Although this relation is not exact
for $\hat{v}^{\mathrm{(C\rho)}}_{ij}[\bar{\rho}(\mathbf{r}_i;\beta)]$
because of the $\beta$-dependence of $\bar{\rho}$,
Eq.~\eqref{eq:DeltaS01_correlation} holds approximately.
Figure~\ref{beta_dep_int_HF+BCS} shows
the $g$ dependence of $-\varDelta \mathcal{S}^{01}(\beta)$
in Eq.~\eqref{eq:Delta_S01}
for $\hat{\mathcal{S}}=\hat{V}^{\mathrm{(C)}}$ and $\hat{V}^{\mathrm{(C\rho)}}$
in $^{34}_{12}$Mg and $^{80,100}_{~~~~40}$Zr.
As $g$ increases,
the behavior of $-\varDelta \mathcal{S}^{01}(\beta)$ drastically changes.
The signs of $-\varDelta \mathcal{S}^{01}(\beta)$ near $\beta=0$,
which is related to $C[\hat{\mathcal{S}}, \hat{J}_y^{\,2}]$
via Eq.~\eqref{eq:DeltaS01_correlation},
changes from positive (negative) to negative (positive)
for $\hat{\mathcal{S}}=\hat{V}^{\mathrm{(C)}}$ ($\hat{V}^{\mathrm{(C\rho)}}$).
The results in Fig. \ref{beta_dep_int_HF+BCS} correspond to
those in Fig. \ref{percentage_HF+BCS}
via the MoI of Eq.~\eqref{eq:J(J+1)_def_moi}.
As a function of $\beta$,
$-\varDelta \mathcal{S}^{01}(\beta)$ for $^{34}_{12}$Mg
changes more slowly than those for $^{80,100}_{~~~~40}$Zr.
This almost flat structure of $-\varDelta \mathcal{S}^{01}(\beta)$
for $^{34}_{12}$Mg
gives rise to the deviation from the $J(J+1)$ rule
and the irregular $J$ ordering in Fig.~\ref{ratio_HFB}.
The results in Fig. \ref{beta_dep_int_HF+BCS} correspond well
to those in Figs. \ref{ratio_divide_HF+BCS} and \ref{percentage_HF+BCS}.

\begin{figure}[]
\centering
\includegraphics[width=10cm,clip]{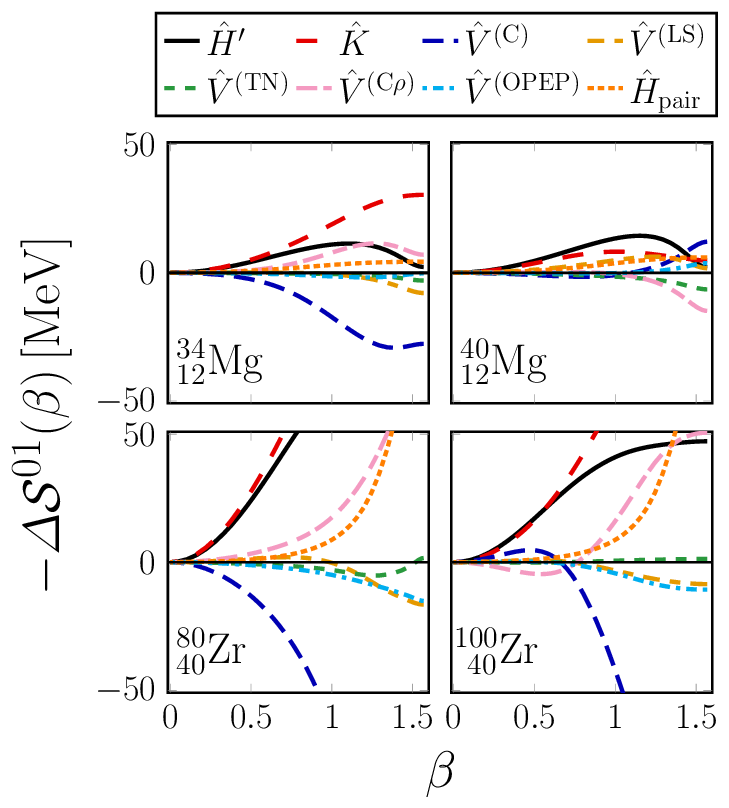}
\caption[]
   {$-\varDelta \mathcal{S}^{01}(\beta)$
     for the constituent terms of the Hamiltonian
     for the HFB solutions
     of $^{34,40}_{~~~12}$Mg and $^{80,100}_{~~~~40}$Zr.}
\label{beta_dep_HFB}
\end{figure}

In Fig.~\ref{beta_dep_HFB},
we show $-\varDelta \mathcal{S}^{01}(\beta)$ for the HFB solutions
of $^{34,40}_{~~~12}$Mg and $^{80,100}_{~~~~40}$Zr.
These results are relevant to those in Fig. \ref{percentage_HFB_pro}.
The $|C[\hat{\mathcal{S}}, \hat{J}_y^{\,2}]|$ values
(see Eq.~\eqref{eq:DeltaS01_correlation})
significantly increase for $\hat{\mathcal{S}}=\hat{K}$, $\hat{V}^{\mathrm{(C)}}$
and $\hat{V}^{\mathrm{(C\rho)}}$ as the mass number increases.
The signs of $C[\hat{\mathcal{S}}, \hat{J}_y^{\,2}]$
for $\hat{H}_{\mathrm{pair}}$ are positive without exceptions.
The signs of $C[\hat{\mathcal{S}}, \hat{J}_y^{\,2}]$
for $\hat{\mathcal{S}}=\hat{V}^{\mathrm{(C)}}$ are negative,
and those for $\hat{V}^{\mathrm{(C\rho)}}$ are positive
for $^{34}_{12}$Mg and $^{80}_{40}$Zr,
which are opposite to the results for the HF solutions
in Fig.~13 of Ref.~\cite{AN22}.
The pair correlations could change
the signs of $C[\hat{\mathcal{S}}, \hat{J}_y^{\,2}]$
for $\hat{\mathcal{S}}=\hat{V}^{\mathrm{(C)}}$ and $\hat{V}^{\mathrm{(C\rho)}}$,
leading to the results in Fig. \ref{percentage_HFB_pro}.
Even though the HFB solution of $^{100}_{~40}$Zr has the pair correlations,
$C[\hat{\mathcal{S}}, \hat{J}_y^{\,2}]$ is positive (negative)
for $\hat{\mathcal{S}}=\hat{V}^{\mathrm{(C)}}$ ($\hat{V}^{\mathrm{(C\rho)}}$).
The flat structure of $-\varDelta \mathcal{S}^{01}(\beta)$
for $\hat{\mathcal{S}}=\hat{V}^{\mathrm{(LS)}}$ in $^{34}_{12}$Mg
and $\hat{V}^{\mathrm{(C\rho)}}$ in $^{40}_{12}$Mg
corresponds to the deviation from $J(J+1)/6$
and the irregular $J$ ordering in Fig. \ref{ratio_HFB}.
The results in Fig.~\ref{beta_dep_HFB} well account for
those in Figs. \ref{ratio_HFB} and \ref{percentage_HFB_pro}.

\subsection{Degree of proximity for nucleons
  associated with nucleonic interaction}
\label{sec:DoP_nucl_int}

We have found that the pairing greatly influences
the contribution of $\hat{V}^{(\mathrm{C})}$ and $\hat{V}^{(\mathrm{C}\rho)}$
to the rotational energy.
In the preceding subsection,
their contributions have been analyzed
in terms of $-\varDelta \mathcal{S}^{01}(\beta)$.
To investigate what governs the contributions
of $\hat{V}^{(\mathrm{C})}$ and $\hat{V}^{(\mathrm{C}\rho)}$ further,
we calculate the DoP $\ev*{\hat{D}}$ by using the AMP,
which has been defined in Sec.~\ref{subsec:DoP}.

\begin{figure}[]
\centering
\includegraphics[width=13cm,clip]{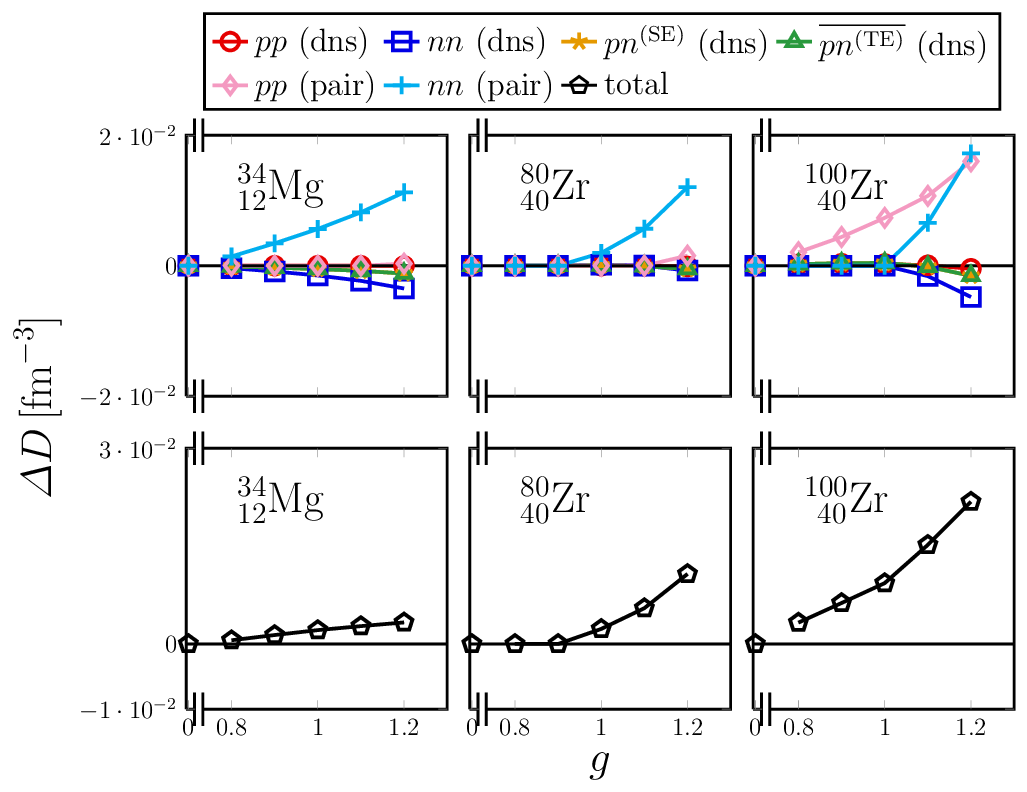}
\caption[]
{
  $\varDelta D$ for the HF+BCS solutions
  of $^{34}_{12}$Mg and $^{80,100}_{~~~~40}$Zr.
  The black pentagons in the lower panels are the total values of $\varDelta D$.
  The separated values of $\varDelta D$ into the individual isospin components
  are also shown;
  the $pp$ (red circles),  $nn$ (blue squares),
  and $pn$ correlations (yellow stars) of the SE channel
  in $\ev*{\hat{D}_{\mathrm{dns}}}$,
  where $p$ ($n$) stands for protons (neutrons).
  The arithmetic average of the spin components is plotted
  for the $\varDelta D$ values of the TE channel (green triangles).
  The $pp$ and $nn$ correlations of the pairing channel
  in $\ev*{\hat{D}_{\mathrm{pair}}}$ in Eq.~\eqref{eq:D_dns_pair}
  are represented as pink diamonds and sky-blue pluses, respectively.
}
\label{delta_int_sa_HF+BCS}
\end{figure}

Denoting the increment of $\ev*{\hat{D}}$ by $\varDelta D$ as
\begin{equation}\label{eq:def_delta_D_g}
\varDelta D:=
\ev*{\hat{D}}{\Phi_0}_g-\ev*{\hat{D}}{\Phi_0}_{g=0},
\end{equation}
the $g$ dependence of $\varDelta D$ is shown for the HF+BCS solutions
of $^{34}_{12}$Mg and $^{80,100}_{~~~~40}$Zr
in Fig.~\ref{delta_int_sa_HF+BCS}.
The $\ev*{\hat{D}}{\Phi_0}_{g=0}$ values
for $^{34}_{12}$Mg, $^{80}_{40}$Zr and $^{100}_{~40}$Zr
are $1.22$, $3.43$ and $4.28$, respectively.
As expected, $\varDelta D$ increases for increasing $g$.
Analogously to Eq.~\eqref{eq:H_with_g},
we separate the DoP $\ev*{\hat{D}}$ as
\begin{equation}\label{eq:D_dns_pair}
\ev*{\hat{D}}=\ev*{\hat{D}_{\mathrm{dns}}}+\ev*{\hat{D}_{\mathrm{pair}}}.
\end{equation}
Owing to the locality of the operator $\hat{D}$,
$\ev*{\hat{D}_{\mathrm{dns}}}$ consists of the SE (viz. $T=1$)
and the TE (viz. $T=0$) channels,
while $\ev*{\hat{D}_{\mathrm{pair}}}$ has only the SE channel.
The $\varDelta D$ values for the individual isospin components,
$pp$, $nn$, $pn$ with $T=1$ and $0$, are also shown.
For the TE channel,
we take the arithmetic average of the three components of the spin-triplet.
The contributions of $\ev*{\hat{D}_{\mathrm{pair}}}{\Phi_0}_g$ to $\varDelta D$
tend to be larger than those of $\ev*{\hat{D}_{\mathrm{dns}}}{\Phi_0}_g$.

\begin{figure}[]
\centering
\includegraphics[width=15cm,clip]{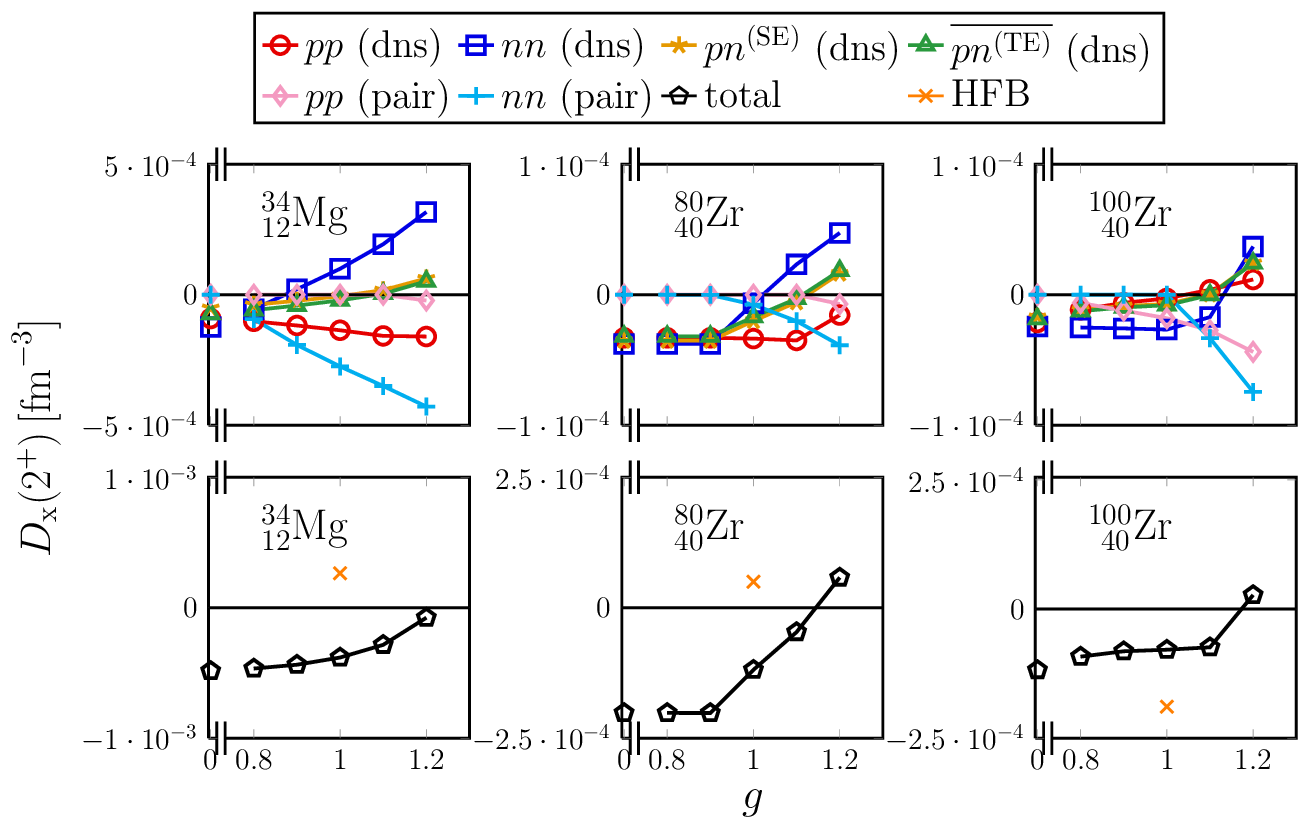}
\caption[]
{
  $D_{\mathrm{x}}(2^+)$ in the HF+BCS solutions
  of $^{34}_{12}$Mg and $^{80,100}_{~~~~40}$Zr.
  See Fig.~\ref{delta_int_sa_HF+BCS} for conventions.
  The orange crosses are the total values of $D_{\mathrm{x}}(2^+)$
  for the HFB solutions.
}
\label{delta_int_exc_HF+BCS}
\end{figure}

We have calculated $\ev*{\hat{D}}{J}$, the DoP at angular-momentum eigenstates,
for the HF+BCS and HFB solutions of $^{34}_{12}$Mg and $^{80,100}_{~~~~40}$Zr.
The $g$ dependence of $\ev*{\hat{D}}{J}_{g}-\ev*{\hat{D}}{J}_{g=0}$
behaves similarly to that of $\varDelta D$ in Fig. \ref{delta_int_sa_HF+BCS}.
We then define $D_{\mathrm{x}}(J^+)$ as
\begin{equation}\label{eq:delta_x}
D_{\mathrm{x}}(J^+)
:= \ev*{\hat{D}}{J} - \ev*{\hat{D}}{0},
\end{equation}
which corresponds with Eq.~\eqref{eq:S_x} for $\hat{\mathcal{S}}=\hat{D}$.
As the intrinsic state $\ket{\Phi_0}$ is identical
among $\ket{J}$ ($J=0,2,4,\cdots$),
$D_{\mathrm{x}}(J^+)$ are small, but do not vanish.
The $g$ dependence of $D_{\mathrm{x}}(2^+)$ for the HF+BCS solutions
of $^{34}_{12}$Mg and $^{80,100}_{~~~~40}$Zr are exhibited
in Fig.~\ref{delta_int_exc_HF+BCS}.
The values of $D_{\mathrm{x}}(2^+)$ for the HFB solutions (with $g=1$)
are also shown for reference.
In $^{80}_{40}$Zr,
$D_{\mathrm{x}}(2^+)$ for the HFB solution is close
to those for the HF+BCS solutions at $g\approx 1.2$.
No $g$ values match the HFB results in $^{34}_{12}$Mg and $^{100}_{~40}$Zr,
suggesting sizable influence of the pairing on the HF configurations
as discussed in Subsec.~\ref{subsec:deformation}.

For the HF (viz. $g=0$) solutions,
the $D_{\mathrm{x}}(2^+)$ values are negative.
This means that the DoP $\ev*{\hat{D}}{J}$ decreases
and the nucleons slightly spread as $J$ goes up.
As $g$ increases from the critical points, $D_{\mathrm{x}}(2^+)$ increases.
The signs of $D_{\mathrm{x}}(2^+)$ change between $g=1.1$ and $1.2$.
Thus, although the nucleons spread with increasing $J$ at the HF level,
the pair correlations reduce and finally invert the effect.
The $D_{\mathrm{x}}(2^+)$ values are decomposed
into the individual isospin components.
As $g$ increases,
contributions of $\ev*{\hat{D}_{\mathrm{dns}}}$ to $D_{\mathrm{x}}(2^+)$
also increase irrespective of the four isospin components,
while those of $\ev*{\hat{D}_{\mathrm{pair}}}$ decrease from zero.
We should note that there will be counter effects in actual nuclei
that are not included in the pure rotational energy.
The intrinsic state may gradually stretch with increasing $J$,
as handled in the cranking model~\cite{PT62,Ka68,RS80}
and the VAP schemes~\cite{RS80}.

These results correlate well
with $\mathcal{S}_{\mathrm{x}}(2^+)/E_{\mathrm{x}}(2^+)$
in Fig.~\ref{percentage_HF+BCS},
particularly for $\hat{\mathcal{S}}=\hat{V}^{\mathrm{(C)}}$
and $\hat{V}^{\mathrm{(C\rho)}}$.
This correlation seems to reflect the short-range nature
of the nucleonic interaction.
It is interpreted that the contributions of the individual components
of the interaction to the rotational energy
are governed by the spatial proximity among constituent nucleons
measured by the DoP.
Although the proximity changes with $J$ only by a small fraction,
its effects on the pure rotational energy are significant.
This argument is supported by comparing the fraction
with those for $\hat{V}^{\mathrm{(C)}}$ and $\hat{V}^{\mathrm{(C\rho)}}$,
\textit{i.e.}, the percentage of the contribution to the rotational energy
in the whole $\hat{V}^{\mathrm{(C)}}$ and $\hat{V}^{\mathrm{(C\rho)}}$.
If we estimate it
through $\mathcal{S}_{\mathrm{x}}(2^+)/\ev*{\hat{\mathcal{S}}}{0}$
and take $^{80}_{40}$Zr as an example,
the value is $3.1\times 10^{-5}$ ($1.8\times 10^{-5}$)
for $\hat{\mathcal{S}}=\hat{V}^{\mathrm{(C)}}$
($\hat{\mathcal{S}}=\hat{V}^{\mathrm{(C\rho)}}$).
The corresponding value
for the DoP ($\hat{\mathcal{S}}=\hat{D}$) is $1.5\times 10^{-5}$.
The ratios with the same order of magnitude
are compatible with the interpretation
that the DoP plays an essential role
in the contribution of the nucleonic interaction to the rotational energy.
The consequence also applies to the HF results reported in Ref.~\cite{AN22},
and to the effects of the pairing.
When the DoP diminishes (\textit{i.e.}, the constituent nucleons tend to spread)
for increasing $J$,
the attractive (repulsive) forces
like $\hat{V}^{\mathrm{(C)}}$ ($\hat{V}^{\mathrm{(C\rho)}}$)
increase (decrease) the rotational energy,
and \textit{vice versa},
mediated by the overlap functions.

\section{Conclusion}

The pure rotational energy of nuclei,
\textit{i.e.}, the rotational energy for a fixed intrinsic state,
has extensively been analyzed
by applying the AMP to the self-consistent axial-MF solutions
with the semi-realistic effective Hamiltonian M3Y-P6.
The contributions of the constituent terms of the Hamiltonian
to the total rotational energies have been inspected,
focusing on effects of the pairing.

Due to the pair correlations,
the compositions of the pure rotational energy drastically change,
sometimes inverting their signs,
and depend strongly on nuclides even for the well-deformed nuclei.
When the pairing becomes stronger,
the contributions of the kinetic energies increase.
Those of the attractive (repulsive) forces decrease (increase),
and their signs could invert.

The degree of proximity (DoP) between nucleons slightly
depends on the angular momentum $J$,
and could account for the effects of nucleonic interactions
on the rotational energy.
The nucleons slightly spread as $J$ increases at the HF level,
while the pair correlations can reduce or invert the effect.
It is concluded that the rotational energy of nuclei is carried
by the kinetic energy in its major part,
but is contributed by the nucleonic interaction as well,
sensitively reflecting the DoP, \textit{i.e.}, the degree
how frequently two constituent nucleons come close.
This spatial correlation accounts for
the stable composition of the pure rotational energy
in the HF states of well-deformed medium-to-heavy nuclei
found in Ref.~\cite{AN22},
the deviation in light nuclei and weakly-deformed states,
and the disarrangement due to the pairing.
For actual nuclei,
the intrinsic state varies as $J$ increases,
and this effect influences the DoP and the rotational energy,
which is ignored in the present study and left for future works.
Still, even when the intrinsic state depends on $J$,
the role of the DoP discovered here
will give us an insight into the effects of the interaction
on the rotational energy of nuclei.

\section*{Acknowledgments}

The authors are grateful to H.~Kurasawa, K.~Yoshida and K.~Washiyama
for discussions.
A part of this work was performed
under the long-term international workshop
on ``Mean-field and Cluster Dynamics in Nuclear Systems (MCD2022)'',
sponsored by the Yukawa International Program for Quark-Hadron Sciences
and held at Yukawa Institute for Theoretical Physics (YITP), Kyoto University,
Japan.
This research was partly supported by the research assistant program at Chiba University.
Numerical calculations were carried out on Yukawa-21 at YITP,
Oakforest PACS at Center for Computational Sciences, University of Tsukuba
under the Multidisciplinary Cooperative Research Program,
and HITACHI SR24000 at the Institute of Management and Information Technologies, Chiba University.


\end{document}